\documentclass[%
reprint,
superscriptaddress,
amsmath,amssymb,
aps,
prl
]{revtex4-2}
\usepackage[T1]{fontenc}
\usepackage{lmodern}
\usepackage[utf8]{inputenc}
\usepackage[english]{babel}
\usepackage{graphicx}
\usepackage{dcolumn}
\usepackage{bm}     
\usepackage{hyperref}
\usepackage{xcolor}
\usepackage{soul}
\usepackage{braket}
\usepackage{amsmath} 
\usepackage{upgreek}
\hypersetup{colorlinks=true,citecolor={blue},linkcolor={blue},urlcolor={blue}}

\usepackage{lineno}
\usepackage{fancyhdr} 
\begin{document}
    \fancyhf{}
    \cfoot{\thepage}
    \pagestyle{fancy}  
	\pagenumbering{arabic}
    \makeatletter
    \let\ps@titlepage\ps@plain
    \makeatother
	\title{In-situ tunneling control in photonic potentials by Rashba-Dresselhaus spin-orbit~coupling}
	
	\author{Rafa\l{}\,Mirek}
	\altaffiliation[]{These authors contributed equally to this letter.}
	\affiliation{IBM Research Europe - Zurich, S{\"a}umerstrasse 4, R{\"u}schlikon, Switzerland}
	\email[]{rafal.mirek@ibm.com}
	
	\author{Pavel\,Kokhanchik}
	\altaffiliation[]{These authors contributed equally to this letter.}
	\affiliation{Universit\'e Clermont Auvergne, Clermont Auvergne INP, CNRS, Institut Pascal, F-63000 Clermont-Ferrand, France}
	
	\author{Darius\,Urbonas}
	\affiliation{IBM Research Europe - Zurich, S{\"a}umerstrasse 4, R{\"u}schlikon, Switzerland}
	
	\author{Ioannis\,Georgakilas}
	\affiliation{IBM Research Europe - Zurich, S{\"a}umerstrasse 4, R{\"u}schlikon, Switzerland}
	
	\author{Marcin\,Muszy\'nski}
	\affiliation{Institute of Experimental Physics, Faculty of Physics, University of Warsaw, Poland}
	
	\author{Piotr\,Kapu\'sci\'nski}
	\affiliation{Institute of Experimental Physics, Faculty of Physics, University of Warsaw, Poland}
	
	\author{Przemys\l{}aw\, Oliwa}
	\affiliation{Institute of Experimental Physics, Faculty of Physics, University of Warsaw, Poland}
	
	\author{Barbara\,Pi\k{e}tka}
	\affiliation{Institute of Experimental Physics, Faculty of Physics, University of Warsaw, Poland}
	
	\author{Jacek\,Szczytko}
	\affiliation{Institute of Experimental Physics, Faculty of Physics, University of Warsaw, Poland}
	
	\author{Michael\,Forster}
	\affiliation{Macromolecular Chemistry Group and Wuppertal Center for Smart Materials \& Systems (CM@S), Bergische Universit\"{a}t Wuppertal, Gauss Strasse 20, 42119 Wuppertal, Germany}
	
	\author{Ullrich\,Scherf}
	\affiliation{Macromolecular Chemistry Group and Wuppertal Center for Smart Materials \& Systems (CM@S), Bergische Universit\"{a}t Wuppertal, Gauss Strasse 20, 42119 Wuppertal, Germany}
	
	\author{Przemys\l{}aw\,Morawiak}
	\affiliation{Institute of Applied Physics, Military University of Technology, Warsaw, Poland}
	
	\author{Wiktor\,Piecek}
	\affiliation{Institute of Applied Physics, Military University of Technology, Warsaw, Poland}
	
	\author{Przemys\l{}aw\,Kula}
	\affiliation{Institute of Chemistry, Military University of Technology, Warsaw, Poland}
	
	\author{Dmitry\,Solnyshkov}
	\affiliation{Universit\'e Clermont Auvergne, Clermont Auvergne INP, CNRS, Institut Pascal, F-63000 Clermont-Ferrand, France}
	\affiliation{Institut Universitaire de France (IUF), 75231 Paris, France}
	
	\author{Guillaume\,Malpuech}
	\affiliation{Universit\'e Clermont Auvergne, Clermont Auvergne INP, CNRS, Institut Pascal, F-63000 Clermont-Ferrand, France}
	
	\author{Rainer\,F.\,Mahrt}
	\affiliation{IBM Research Europe - Zurich, S{\"a}umerstrasse 4, R{\"u}schlikon, Switzerland}
	
	\author{Thilo\,St{\"o}ferle}
	\affiliation{IBM Research Europe - Zurich, S{\"a}umerstrasse 4, R{\"u}schlikon, Switzerland}
	
	\begin{abstract}
		
		The tunability of individual coupling amplitudes in photonic lattices is highly desirable for photonic Hamiltonian engineering and for studying topological transitions \textit{in situ}. In this work, we demonstrate the tunneling control between individual lattice sites patterned inside an optical microcavity. The tuning is achieved by applying a voltage to a liquid crystal microcavity possessing photonic Rashba-Dresselhaus spin-orbit coupling. This type of spin-orbit coupling emerges due to the high birefringence of the liquid crystal material and constitutes an artificial gauge field for photons. The proposed technique can be combined with strong-light matter coupling and non-Hermitian physics already established in liquid crystal microcavities. 
		
	\end{abstract}
	
	\maketitle

	\section{Introduction}
	Topological photonics~\cite{lu2014topological,price2022roadmap}, which has gained enormous interest in recent decades, involves numerous effects like the creation of one-way edge states~\cite{Haldane2008,wang2009observation,nalitov2015Z,klembt2018exciton,slobozhanyuk2017three}, structures for topological lasing~\cite{Solnyshkov2016kzm,st2017lasing,bahari2017nonreciprocal,bandres2018topological,choi2021room}, tuning of the topology~\cite{kudyshev2019tuning}, guiding of light~\cite{lumer2019light,chalabi2020guiding,vakulenko2023adiabatic}, non-Abelian physics~\cite{terccas2014non,chen2019non,polimeno2021experimental,yan2023non,yang2024non} (including the non-Abelian Hofstadter model~\cite{yang2020non}), Klein tunneling in topological structures~\cite{ni2018spin}, and chiral zero mode formation at Weyl points~\cite{jia2019observation} that can be described in terms of artificial gauge fields~\cite{aidelsburger2018artificial,chen2019non}, in real~\cite{aidelsburger2018artificial} and reciprocal space~\cite{Price2014}. In the frameworks of Floquet engineering~\cite{rechtsman2013photonic,hey2018advances,yang2020photonic}, synthetic dimensions~\cite{ozawa2016synthetic,lustig2019photonic,dutt2020single}, and spin-orbit coupling (SOC)~\cite{terccas2014non,nalitov2015spin,whittaker2021optical,polimeno2021experimental}, often creative but sophisticated fabrication geometries are used to build an artificial gauge field. Various realizations of artificial gauge fields are based on lattice coupling control techniques, e.g., amplitude control techniques, as in the case of Landau levels observed in strained lattices~\cite{rechtsman2013strain,jamadi2020direct,barczyk2024observation}, or phase control techniques, allowing to realize the Harper-Hofstadter model in coupled resonator arrays~\cite{hafezi2011robust,umucalilar2011artificial} and topological insulator lasers \cite{harari2018topological,bandres2018topological}. Most of these techniques do not allow controlling the couplings \textit{in situ}, that is, to tune them during the experiment, which limits the versatility and the scope of possible applications.
	
	Lately, a new type of photonic SOC called Rashba-Dresselhaus spin-orbit coupling (RDSOC) was developed in highly birefringent optical microcavities arising from the coupling of modes of orthogonal polarization and opposite parity~\cite{rechcinska2019engineering,ren2021nontrivial}. 
	%It can be interpreted as an emergent optical activity.
	Similar SOC has also been obtained in microwave metamaterials~\cite{liu2019polarization}. Since then, photonic RDSOC in microcavities has been studied extensively and various effects such as persistent spin helix~\cite{krol2021realizing, muszynski2022realizing}, tunable Berry curvature~\cite{polimeno2021tuning,lempicka2022electrically}, polariton lasing~\cite{li2022manipulating,long2022helical}, spin Hall effect~\cite{liang2024polariton}, and polariton striped phase~\cite{muszynski2024observation}, have been demonstrated. 
	%In liquid crystal (LC) microcavities, the RDSOC can be switched on and off by applying a voltage or tuned continuously in space by means of a rubbing layer. 
	In a bare liquid crystal (LC) microcavity without any lattice, the RDSOC can already be described as a gauge potential of Abelian~\cite{rechcinska2019engineering,ren2021nontrivial} or non-Abelian nature~\cite{jin20062,polimeno2021experimental}, depending on the presence of a coupling between circular components.
	%This SOC is tunable, continuously in space, and itself an artificial gauge field in a bare LC microcavity without any lattice~\cite{jin20062}. 
	This makes RDSOC more similar to gauge fields for electrons~\cite{bernevig2006exact,koralek2009emergence} than to other synthetic gauge fields in photonics, which are usually achieved by lattice coupling engineering \cite{tang2022topological}. Therefore, we inverse the paradigm and study how the photonic artificial gauge field in the form of RDSOC affects the lattice couplings, providing a useful tool for fundamental and applied studies, because the variation of the couplings allows obtaining different emergent Hamiltonians and tuning them across transitions. In a previous work by some of us, two different regimes of coupling control, i.e., amplitude and phase control, have already been demonstrated theoretically~\cite{kokhanchik2022modulated}. It was shown that the former allows to realize a Su-Schrieffer-Heeger (SSH) chain with a topological phase transition driven by a voltage applied to the microcavity, while the latter provides a way to implement the spinful Harper-Hofstadter model. However, an experimental implementation with a demonstration of \textit{in situ} coupling control has been missing so far. 
	
	\begin{figure*}
		\centering
		\includegraphics[width=0.49\linewidth]{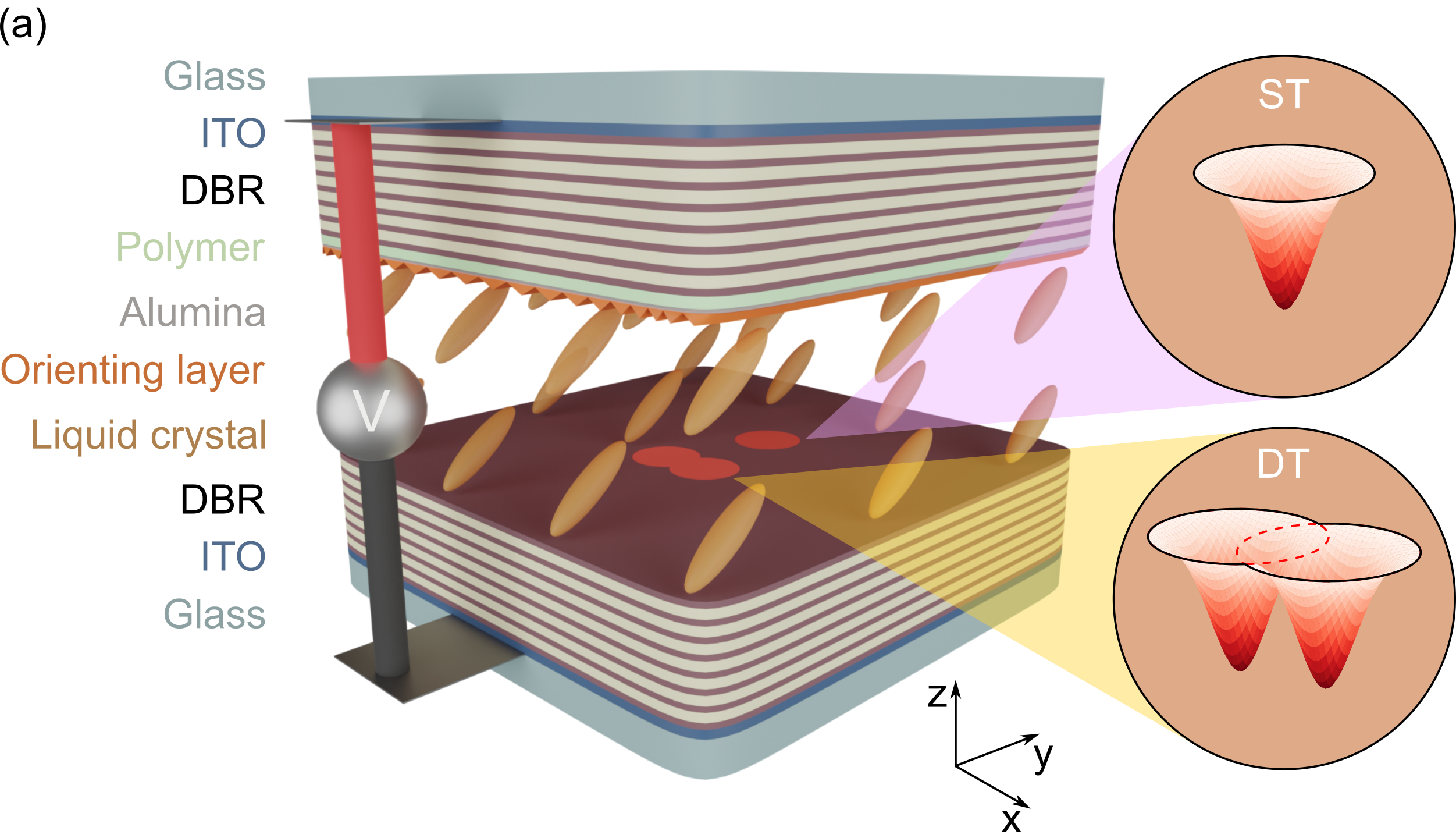}
		\includegraphics[width=0.49\linewidth]{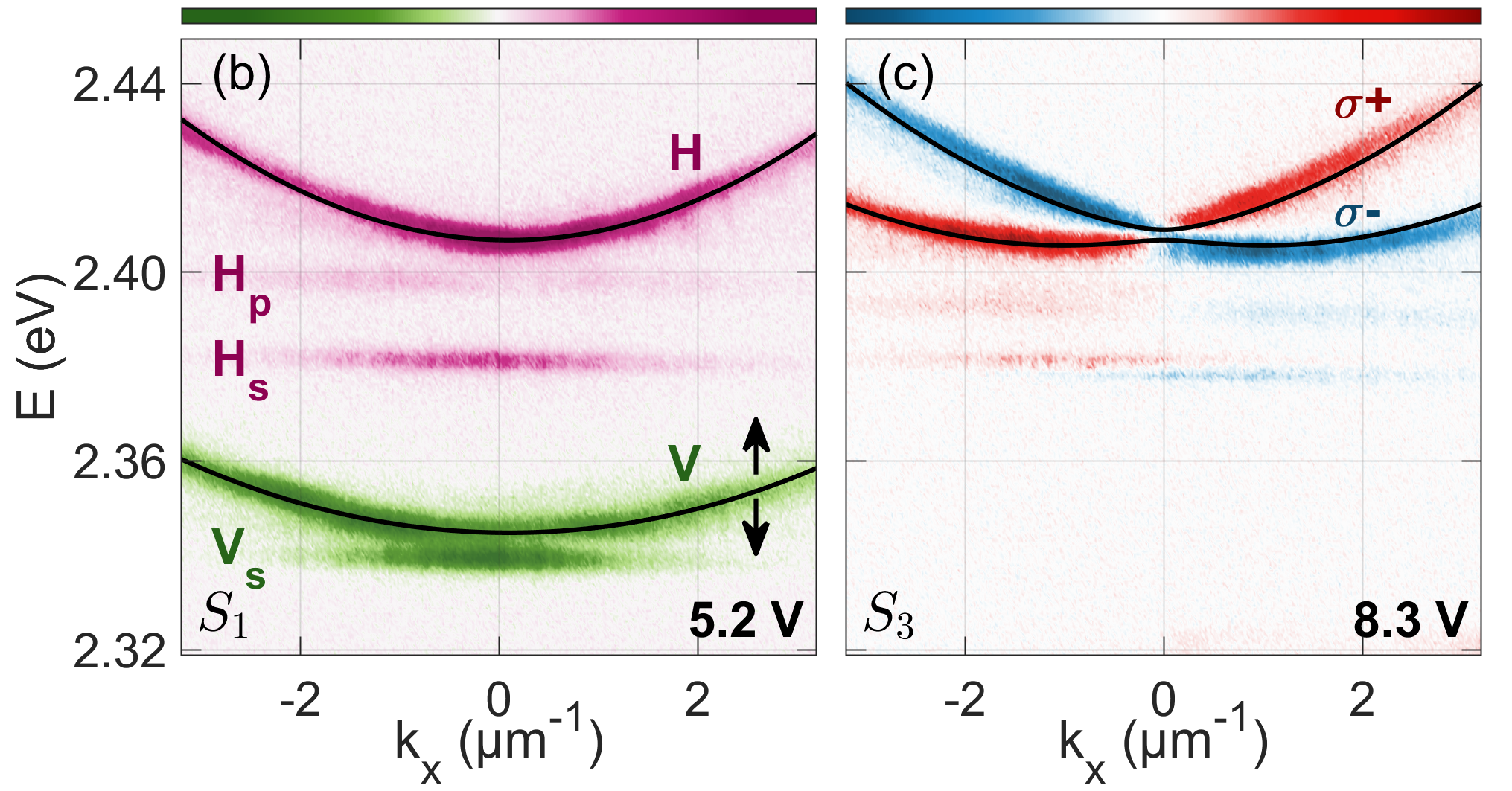}
		\caption{(a) Scheme of LC microcavity with ST and DT potentials; experimentally measured dispersion of ST for (b) 5.2~V and (c) 8.3~V shown as $S_1$ and $S_3$ Stokes parameters, respectively; the data is nonlinearly transformed and saturated in order to enhance the visibility of localized states marked as $V_s$, $H_s$ and $H_p$; the colorbars are in arb. units.; black solid lines show the fit of continuous modes (H and V in (b), $\sigma^+$ and $\sigma^-$ in (c)) by Hamiltonian~\eqref{RDSOC_Ham}; black double-headed arrow depicts the tunability of V mode. See parameters in~\cite{suppl}.}
		\label{fig_1}
	\end{figure*}
	
	In this work, we experimentally demonstrate coupling amplitude control achieved through RDSOC. We start by studying a single Gaussian trap (ST) in an LC microcavity. We demonstrate coupling by RDSOC between orthogonally linearly-polarized localized $s$ and $p$ states of ST. Next, we study a double Gaussian trap (DT) and show how the coupling amplitude between two collinearly polarized localized $s$ states can be controlled through RDSOC. The coupling can be continuously tuned with applied voltage, which is an important step forward even with respect to previous theoretical predictions~\cite{kokhanchik2022modulated}. This kind of tunable coupling control is hard to achieve in any other optical microcavity realization. 
	%It has also not been demonstrated in other photonic platforms.
	An important perspective is opened by the possibility to reduce the coupling to zero and then to invert its sign by increasing the RDSOC. This realizes a topological transition observable as a swap of symmetric and anti-symmetric states in the microcavity spectrum.
	
	\section{Results}
	
	The studied LC microcavity sample is displayed in Fig.~\ref{fig_1}(a). The detailed structure and measurement setup are presented in~\cite{suppl}. To study the optical properties, white light transmission measurements at room temperature have been carried out. The microcavity with distributed Bragg reflectors (DBR) filled with several micrometer thick layer of nematic liquid crystal (LC) naturally possesses longitudinal modes which can be numbered with an integer corresponding to the number of field antinodes. The orientation of the molecular director in the $yz$ plane can be controlled via a voltage applied to the indium tin oxide (ITO) electrodes, resulting in tuning of the effective permittivity $\varepsilon_{yy}$, while $\varepsilon_{xx}$ remains constant. At high voltage, the LC molecules are oriented along the cavity axis $z$, so that H and V modes with the same order number $N$ are in resonance~\cite{rechcinska2019engineering}. Due to the large birefringence of the LC material, while changing voltage, the orthogonally linearly polarized eigenmodes can be significantly detuned, so that the modes of different numbers can approach each other. In the regime where the H-polarized $N$th mode and the V-polarized $(N+1)$th mode are energetically close to each other, the coupling of these bands leads to the RDSOC effect~\cite{rechcinska2019engineering,ren2021nontrivial}. Here, an effective 1D two-band Hamiltonian along the direction perpendicular to the plane of the director rotation, in the basis defined by the two linearly-polarized H and V modes, reads~\cite{rechcinska2019engineering}:
	\begin{equation}
		H\left(k_x\right) = \frac{\hbar^2 k_x^2}{2 M} \mathbb{I} +  \left(\frac{\hbar^2 k_x^2}{2 m} + 2 \alpha k_x\right) \sigma_x + \delta \sigma_1,
		\label{RDSOC_Ham}
	\end{equation}
	with $1/M = (1/m_H + 1/m_V)/2$, $1/m = (1/m_H - 1/m_V)/2$, where $m_H\, (m_V)$ is the mass of the H-(V-)polarized mode, $\alpha$ -- the RDSOC parameter, $\delta$ -- the (linear) detuning between $N$th H and $(N+1)$th V modes, $k_x$ -- the wavevector component along the RDSOC direction. Finally, $\sigma_1 = (\mathbb{I} - \sigma_z)/2$, where $\mathbb{I}$ is the identity matrix, $\sigma_{x,z}$ are the Pauli matrices. $\sigma_1$ is used because the applied voltage shifts only the V-polarized band.
	
	In Fig.~\ref{fig_1}, we show the dispersion measured at the position of an ST for different voltages. At small voltages, the detuning $\delta$ between the two linear polarizations is large, and therefore, the eigenmodes are predominantly linearly polarized. It is clearly visible when plotting the first Stokes parameter (degree of linear polarization) $S_1$ (Fig.~\ref{fig_1}(b)). Along with continuous parabolic cavity modes, well described by the Hamiltonian~\eqref{RDSOC_Ham}, we observe the localized states. Due to the contrast in H and V effective refractive indices, the effective potential depth of the ST is different for H and V. As a result, a single localized state is observed for V and two localized states for H. We call them $s$ and $p$ levels, in analogy with atomic orbitals.
	
	For higher voltages (around 8.3~V), the V mode of $(N+1)$th order approaches the H mode of $N$th order, and the RDSOC contribution $2 \alpha k_x$ starts to dominate over the splitting $\delta$ between the linearly-polarized modes (Fig.~\ref{fig_1}(c)). In this regime, the eigenmodes are mostly circularly polarized and split along $k_x$, which is visible when plotting the third Stokes parameter (degree of circular polarization) $S_3$. Indeed, the $2 \alpha k_x \sigma_x$ term produces an effect similar to the optical activity \cite{ren2021nontrivial}.
	
	\begin{figure}
		\centering
		\includegraphics[width=\linewidth]{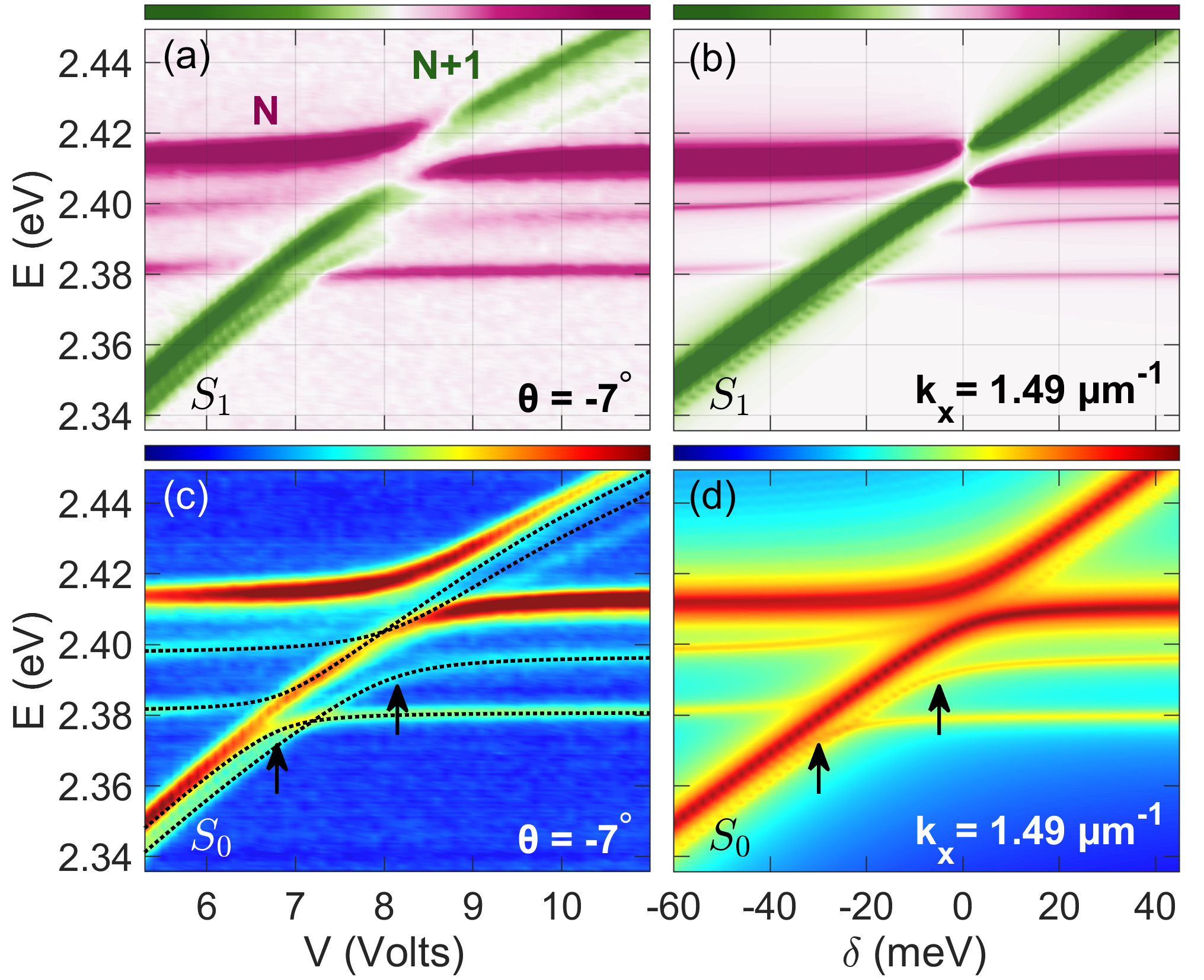}
		\caption{Energy-voltage diagrams showing the near-resonant regime between $N$th H mode and $(N+1)$th V mode in an ST. (a,c) Experimental and (b,d) numerical results represented as (a,b) linear polarization degree $S_1$ and (c,d) total intensity $S_0$; the detection angle of experimental data $\theta=-7^{\circ}$ corresponds to $k_x=1.49$ $\mu m^{-1}$ of the numerical simulation; the data is nonlinearly transformed and saturated in order to enhance visibility of localized states; the colorbars are in arb. units; black arrows in (c,d) mark anticrossing points; black dotted lines in (c) show the energies obtained by diagonalizing tight-binding Hamiltonian~\eqref{Ham_ST}. See simulation parameters and other Stokes components for full 0-11 V voltage range in~\cite{suppl}.}
		\label{fig_2}
	\end{figure}
	
	We measured the dispersion at the position of an ST for the set of voltages between 0~V and 11~V. The results are shown in Fig.~\ref{fig_2}(a) in an energy-voltage plot for a fixed angle of detection of $-7^\circ$ corresponding to $k_x$=1.49\,$\upmu$m$^{-1}$ (which optimizes the simultaneous visibility of both $s$ and $p$ localized states). The tunability of the V mode with voltage is well visible in this type of diagram. The experimental results can be reproduced by numerically solving the stationary Schrödinger equation with the Hamiltonian~\eqref{RDSOC_Ham} as displayed in~Fig.~\ref{fig_2}(b). From the experimentally measured total intensity in Fig.~\ref{fig_2}(c), one can infer that localized states anticross at the positions highlighted by black arrows, corroborated by numerical results (Fig.~\ref{fig_2}(d)). This anticrossing is a manifestation of RDSOC-mediated coupling between particular localized states of H and V. The absolute value of the matrix elements of this coupling can be generally written as $\beta = | \bra{H_l} 2 \alpha k_x \ket{V_{l'}} |$, with $l (l') = \{s,p,...\}$ representing the quantum number of a localized state. Since the RDSOC term is linear in $k_x$, it couples only the localized states of opposite parity ($s$ and $p$ in our case). While the $V_{p}$ state was not visible in Fig.~\ref{fig_1}(b), it appears when H and V are close to resonance, and so it can couple with the other states which are always visible. Therefore, the tight-binding model for localized states of an ST in the basis $(\ket{H_s} \ \ket{H_p} \ \ket{V_s} \ \ket{V_p})^T$ can be written as:
	\begin{equation}
		H_{ST} = 
		\begin{pmatrix}
			E_{H_s} & 0 & 0 & -i \beta_{ST,2} \\
			0 & E_{H_p} & i \beta_{ST,1} & 0 \\
			0 & -i \beta_{ST,1} & E_{V_s} + \delta & 0 \\
			i \beta_{ST,2} & 0 & 0 & E_{V_p} + \delta        
		\end{pmatrix}.
		\label{Ham_ST}
	\end{equation}
	Two different coupling coefficients $\beta_{ST,1(2)}$ give rise to two anticrossing points observed in Fig.~\ref{fig_2}(c,d). By diagonalizing the tight-binding Hamiltonian~\eqref{Ham_ST}, we obtain a good correspondence with experimental results (black dotted lines in Fig.~\ref{fig_2}(c)).
	
	\begin{figure}
		\centering
		\includegraphics[width=\linewidth]{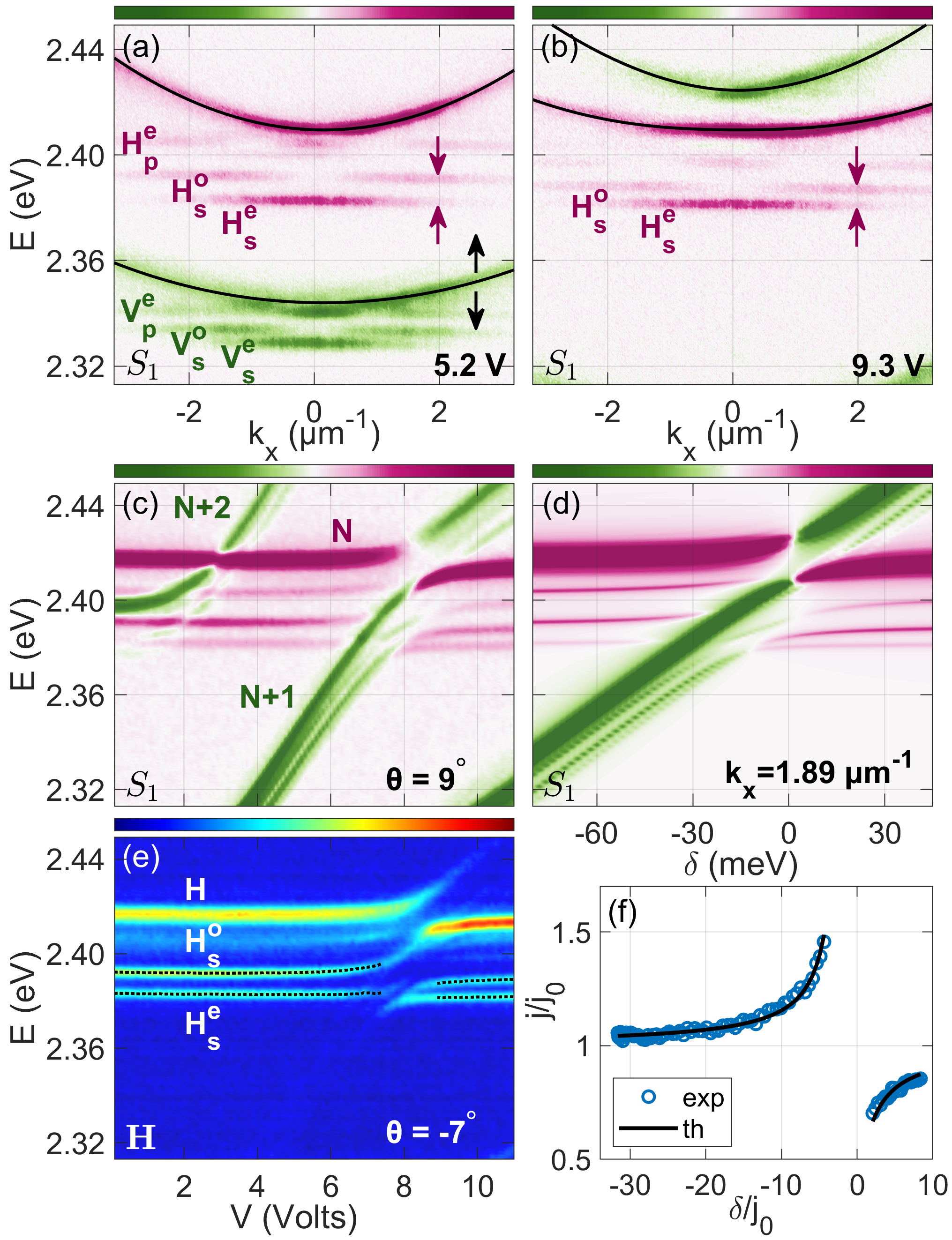}
		\caption{Experimentally measured dispersion of a DT for (a) 5.2~V and (b) 9.3~V shown as $S_1$ Stokes parameter; black solid lines show the fit by Hamiltonian~\eqref{RDSOC_Ham}; black double-headed arrow depicts the tunability of V mode; purple arrows mark the spectral distance between $H_s^e$ and $H_s^o$; (c,e) experimental and (d) numerical results represented as (c,d) $S_1$ linear polarization degree and (e) H intensity; black dotted lines in (e) show the extracted energies of $H_s^e$ and $H_s^o$ localized states; the data in (a-e) is nonlinearly transformed and saturated in order to enhance the visibility of localized states; the colorbars are in arb. units; (f) effective coupling coefficient between $H_s^e$ and $H_s^o$ localized states extracted from the experiment (blue dots) and fitted by $j(\delta)$ (black line). See simulation parameters and other Stokes components in~\cite{suppl}.}
		\label{fig_3}
	\end{figure}
	
	We now turn to the consideration of the DT potential. The dispersions for negative and positive linear detunings are shown in Fig.~\ref{fig_3}(a,b), respectively. Now, each state of an ST is split into two states, since the traps forming the DT are coupled by tunneling. We dub the resulting states "even" and "odd" (superscript notations $e$ and $o$), depending on the symmetry of the corresponding DT eigenstate in real space. For both H and V modes, we observe three localized states. The coupling between different localized states is described by the same matrix elements with $\beta$ introduced above. Note that the energy splitting between $H_s^e$ and $H_s^o$ localized states is different for panels (a) and (b) in Fig.~\ref{fig_3}. Moreover, it changes continuously with voltage (Fig.~\ref{fig_3}(e)). The effect is especially pronounced when approaching the resonance. At the same time, the polarization of the states is mostly conserved, which is confirmed by $S_1$ both in the experiment (Fig.~\ref{fig_3}(c)) and numerical simulation (Fig.~\ref{fig_3}(d)). Since the uncoupled states in both traps are degenerate, the energy splitting between the DT states can be written as $2 j$, where $j$ is a tunneling coefficient. This result can be interpreted as a continuous control of the effective tunneling by the applied voltage due to the RDSOC: $j=j(V)$.
	
	To confirm this hypothesis, we built a 6x6 tight-binding model (see~\cite{suppl}) taking into account all localized states, and then using perturbation theory up to the 2nd order in the limit $|\delta| \gg \max{|\Delta E_{ij}|}, \beta, |j_0|$, where $\Delta E_{ij}$ is a difference between $i$th and $j$th localized states and $j_0$ is the tunneling in the absence of RDSOC, we reduced it to an effective $2\times 2$ Hamiltonian including only $(\ket{H_s^l} \ \ket{H_s^r})^T$ basis states (left and right wells in a DT):
	\begin{equation}
		H_{DT} =
		\begin{pmatrix}
			E_{H_s}'(\delta) & -j(\delta) \\
			-j(\delta) & E_{H_s}'(\delta)
		\end{pmatrix},
		\label{Ham_DT}
	\end{equation}
	with $E_{H_s}'(\delta)$ -- the detuning-dependent modified energy of an ST $H_s$ state (see~\cite{suppl}). Importantly, we obtain an expression for the detuning-dependent effective tunneling:
	\begin{equation}
		j\left(\delta\right) = j_0 -\frac{\beta_{DT,2}^2}{2 \left(E_{V_p^e} - E_{H_s} + \delta\right)}
		\label{efftun}
	\end{equation}
	where $\beta_{DT,2} = |\bra{H_s^o} 2 \alpha k_x \ket{V_p^e}|$, $E_{H_s}$ and $E_{V_p^e}$ are the energies of ST $H_s$ and DT $V_p^e$ states, respectively. The effective tunneling extracted from the experiment is shown as blue points in Fig.~\ref{fig_3}(f), while the fitting of this data with $j(\delta)$ is illustrated by a solid black line.
	We, therefore, conclude that the hyperbolic dependence of the tunneling $j$ on the detuning $\delta$ (directly controlled by the voltage) given by Eq.~\eqref{efftun} is confirmed by the experiment.
	The total change of $j(\delta)$ is from $0.7 j_0$ to $1.45 j_0$, exceeding a factor~$2$. This change is only possible if the absolute value of the RDSOC matrix element $\beta_{DT,2}$ is non-zero. Thus, the RDSOC-mediated coupling of the $s$ and $p$ states plays a crucial role in the control of the coupling between the two traps.
	
	\begin{figure}
		\centering
		\includegraphics[width=\linewidth]{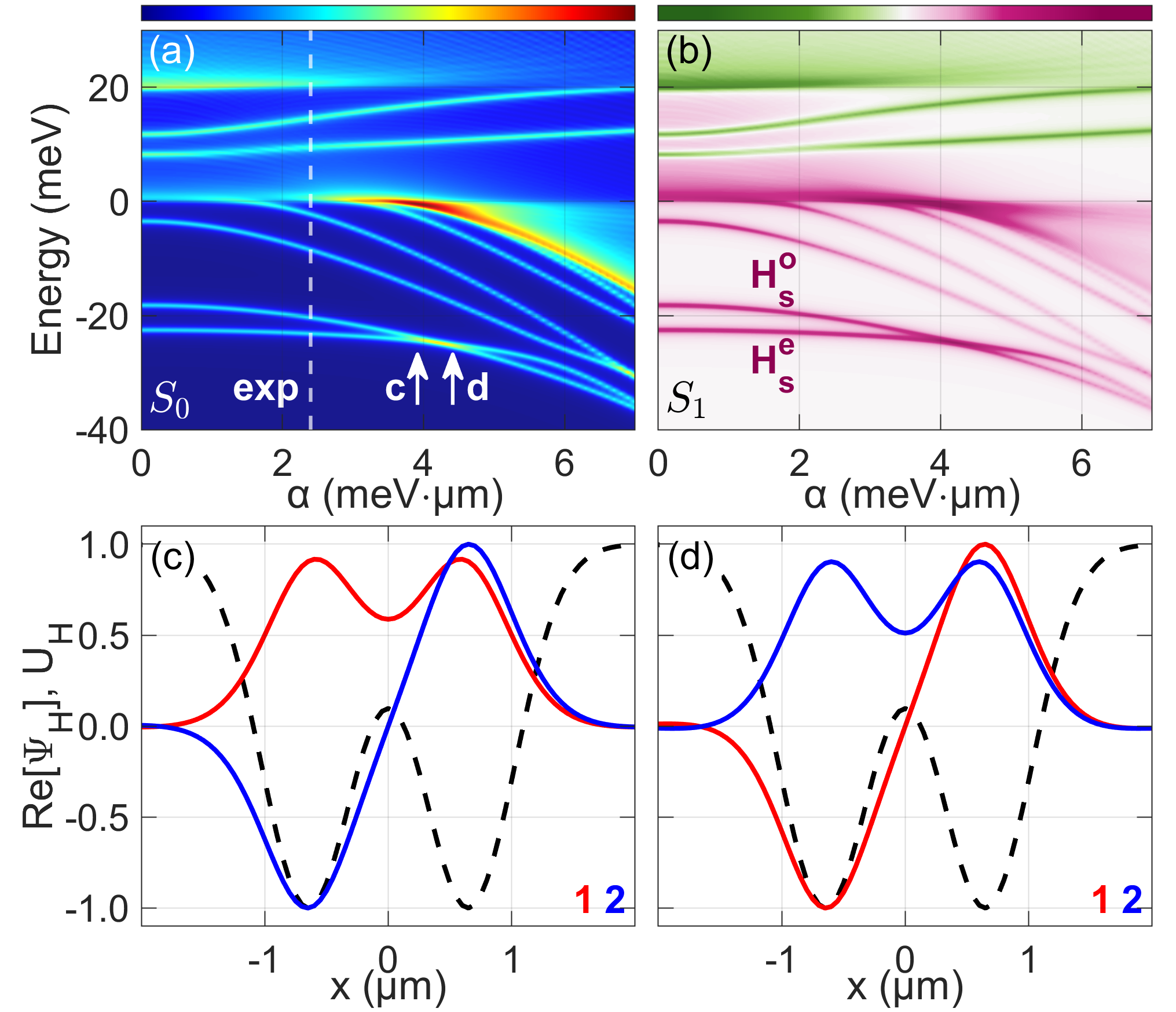}
		\caption{Spectrum of a DT depending on RDSOC parameter $\alpha$ showing a spectral inversion between $H_s^e$ and $H_s^o$ localized states in (a) total intensity $S_0$ and (b) linear polarization degree $S_1$; the colorbars are in arb. units; real part of H-polarized wavefunction component $\rm{Re}[\Psi_H]$ of the lowest (1, red) and the second lowest (2, blue) localized states (c) before and (d) after level inversion; black dashed line shows a potential $U_H$ for H; $\rm{Re}[\Psi_H]$ and $U_H$ units are normalized; white dashed line in (a) shows the experimental value of RDSOC parameter $\alpha$ presented in Fig.~\ref{fig_3}. See parameters in~\cite{suppl}.}
		\label{fig_4}
	\end{figure}
	
	Another remarkable property of the effective coupling $j(\delta)$ is a minus sign allowing, in principle, to suppress the effective coupling down to zero or even to invert its sign. The inversion of the coupling sign is inextricably linked with a topological transition manifesting itself in the experiment as a spectral swap between $e$ and $o$ localized states. In lattices, this exchange of the states of different symmetry is associated with a topological transition \cite{bernevig2006quantum}. In Fig.~\ref{fig_4} we demonstrate this effect in numerical simulations. Figs.~\ref{fig_4}(a,b) show the evolution of the spectrum with increasing RDSOC constant $\alpha$. At $\alpha \approx 4~$meV$\cdot \mu$m, the $H_s^e$ and $H_s^o$ localized states cross in energy. By looking at the structure of the lowest (1) and second lowest (2) states in real space before the crossing (Fig.~\ref{fig_4}(c)) and after the crossing (Fig.~\ref{fig_4}(d)), one can easily notice the inversion between the $e$ and $o$ states. At the same time, the polarization of these states is mostly conserved, as confirmed by the degree of linear polarization $S_1$ (Fig.~\ref{fig_4}(b)).
	
	In our DT experiment (Fig.~\ref{fig_3}), the value of $\alpha$ is 2.4 meV$\cdot \mu$m, which does not allow us to observe this topological transition experimentally (white dashed line in Fig.~\ref{fig_4}(a)). The simplest way to achieve it in future experiments is to utilize LC materials with higher birefringence.
	
	\section{Discussion and Conclusions}
	Along with an abundance of physical phenomena, microcavities allow studying various lattice models, often in the context of topological photonics. Nowadays, the most common techniques for lattice engineering in planar microcavities are micropillar etching~\cite{michaelis2011spatial,solnyshkov2021microcavity}, where the tunneling in the presence of spin-orbit coupling has also been studied~\cite{sala2015spin}, and ballistical coupling~\cite{topfer2021engineering}. Both of these techniques lack a simple tunability for the lattice couplings. This reduces the range of models to be investigated, and, probably more importantly, does not provide dynamical access to the transitions between different topological phases of the model. The technique of coupling amplitude and sign control suggested in the present work, along with the phase control discussed before~\cite{kokhanchik2022modulated}, allows accessing richer topological physics in optical microcavities. This approach can be further extended to the domains of strong light-matter interaction, non-linear physics, or non-Hermitian physics, since all necessary ingredients are available in LC microcavities. Furthermore, the presented coupling control and inversion can be applied in cold atom systems in addition to the existing opportunities~\cite{proukakis2017universal}, since RDSOC is broadly studied there~\cite{lin2011spin}.
	
	In this work, we studied the localized states in LC microcavity in the presence of RDSOC. We demonstrated experimentally the coupling between ST localized states of different parity ($s$ and $p$). We then studied experimentally the DT system and showed how the effective tunneling between the traps can be continuously controlled by the voltage applied to the LC microcavity thanks to the presence of the mentioned RDSOC-mediated $sp$ coupling. Finally, in addition to the coupling amplitude control, we demonstrated theoretically that the coupling sign can be inverted, which shows up as a topological transition in the LC microcavity spectrum.
	
	\section{Acknowledgments}
	We thank the team of the IBM Binnig and Rohrer Nanotechnology Center and D. Caimi for support with the sample fabrication. This work was supported by the European Union’s Horizon 2020 program, through a FET Open research and innovation action under grant agreements No. 964770 (TopoLight) and EU H2020 MSCA-ITN project under grant agreement No. 956071 (AppQInfo). Additional support was provided by the ANR Labex GaNext (ANR-11-LABX-0014), the ANR program "Investissements d'Avenir" through the IDEX-ISITE initiative 16-IDEX-0001 (CAP 20-25), the ANR project MoirePlusPlus, and the ANR project "NEWAVE" (ANR-21-CE24-0019).
	
	\bibliography{main_arXiv}

\begin{thebibliography}{64}%
\makeatletter
\providecommand \@ifxundefined [1]{%
 \@ifx{#1\undefined}
}%
\providecommand \@ifnum [1]{%
 \ifnum #1\expandafter \@firstoftwo
 \else \expandafter \@secondoftwo
 \fi
}%
\providecommand \@ifx [1]{%
 \ifx #1\expandafter \@firstoftwo
 \else \expandafter \@secondoftwo
 \fi
}%
\providecommand \natexlab [1]{#1}%
\providecommand \enquote  [1]{``#1''}%
\providecommand \bibnamefont  [1]{#1}%
\providecommand \bibfnamefont [1]{#1}%
\providecommand \citenamefont [1]{#1}%
\providecommand \href@noop [0]{\@secondoftwo}%
\providecommand \href [0]{\begingroup \@sanitize@url \@href}%
\providecommand \@href[1]{\@@startlink{#1}\@@href}%
\providecommand \@@href[1]{\endgroup#1\@@endlink}%
\providecommand \@sanitize@url [0]{\catcode `\\12\catcode `\$12\catcode
  `\&12\catcode `\#12\catcode `\^12\catcode `\_12\catcode `\%12\relax}%
\providecommand \@@startlink[1]{}%
\providecommand \@@endlink[0]{}%
\providecommand \url  [0]{\begingroup\@sanitize@url \@url }%
\providecommand \@url [1]{\endgroup\@href {#1}{\urlprefix }}%
\providecommand \urlprefix  [0]{URL }%
\providecommand \Eprint [0]{\href }%
\providecommand \doibase [0]{https://doi.org/}%
\providecommand \selectlanguage [0]{\@gobble}%
\providecommand \bibinfo  [0]{\@secondoftwo}%
\providecommand \bibfield  [0]{\@secondoftwo}%
\providecommand \translation [1]{[#1]}%
\providecommand \BibitemOpen [0]{}%
\providecommand \bibitemStop [0]{}%
\providecommand \bibitemNoStop [0]{.\EOS\space}%
\providecommand \EOS [0]{\spacefactor3000\relax}%
\providecommand \BibitemShut  [1]{\csname bibitem#1\endcsname}%
\let\auto@bib@innerbib\@empty
%</preamble>
\bibitem [{\citenamefont {Lu}\ \emph {et~al.}(2014)\citenamefont {Lu},
  \citenamefont {Joannopoulos},\ and\ \citenamefont
  {Solja{\v{c}}i{\'c}}}]{lu2014topological}%
  \BibitemOpen
  \bibfield  {author} {\bibinfo {author} {\bibfnamefont {L.}~\bibnamefont
  {Lu}}, \bibinfo {author} {\bibfnamefont {J.~D.}\ \bibnamefont
  {Joannopoulos}},\ and\ \bibinfo {author} {\bibfnamefont {M.}~\bibnamefont
  {Solja{\v{c}}i{\'c}}},\ }\href@noop {} {\bibfield  {journal} {\bibinfo
  {journal} {Nature Photonics}\ }\textbf {\bibinfo {volume} {8}},\ \bibinfo
  {pages} {821} (\bibinfo {year} {2014})}\BibitemShut {NoStop}%
\bibitem [{\citenamefont {Price}\ \emph {et~al.}(2022)\citenamefont {Price},
  \citenamefont {Chong}, \citenamefont {Khanikaev}, \citenamefont {Schomerus},
  \citenamefont {Maczewsky}, \citenamefont {Kremer}, \citenamefont {Heinrich},
  \citenamefont {Szameit}, \citenamefont {Zilberberg}, \citenamefont {Yang}
  \emph {et~al.}}]{price2022roadmap}%
  \BibitemOpen
  \bibfield  {author} {\bibinfo {author} {\bibfnamefont {H.}~\bibnamefont
  {Price}}, \bibinfo {author} {\bibfnamefont {Y.}~\bibnamefont {Chong}},
  \bibinfo {author} {\bibfnamefont {A.}~\bibnamefont {Khanikaev}}, \bibinfo
  {author} {\bibfnamefont {H.}~\bibnamefont {Schomerus}}, \bibinfo {author}
  {\bibfnamefont {L.~J.}\ \bibnamefont {Maczewsky}}, \bibinfo {author}
  {\bibfnamefont {M.}~\bibnamefont {Kremer}}, \bibinfo {author} {\bibfnamefont
  {M.}~\bibnamefont {Heinrich}}, \bibinfo {author} {\bibfnamefont
  {A.}~\bibnamefont {Szameit}}, \bibinfo {author} {\bibfnamefont
  {O.}~\bibnamefont {Zilberberg}}, \bibinfo {author} {\bibfnamefont
  {Y.}~\bibnamefont {Yang}}, \emph {et~al.},\ }\href@noop {} {\bibfield
  {journal} {\bibinfo  {journal} {Journal of Physics: Photonics}\ }\textbf
  {\bibinfo {volume} {4}},\ \bibinfo {pages} {032501} (\bibinfo {year}
  {2022})}\BibitemShut {NoStop}%
\bibitem [{\citenamefont {Haldane}\ and\ \citenamefont
  {Raghu}(2008)}]{Haldane2008}%
  \BibitemOpen
  \bibfield  {author} {\bibinfo {author} {\bibfnamefont {F.~D.~M.}\
  \bibnamefont {Haldane}}\ and\ \bibinfo {author} {\bibfnamefont
  {S.}~\bibnamefont {Raghu}},\ }\href@noop {} {\bibfield  {journal} {\bibinfo
  {journal} {Physical Review Letters}\ }\textbf {\bibinfo {volume} {100}},\
  \bibinfo {pages} {013904} (\bibinfo {year} {2008})}\BibitemShut {NoStop}%
\bibitem [{\citenamefont {Wang}\ \emph {et~al.}(2009)\citenamefont {Wang},
  \citenamefont {Chong}, \citenamefont {Joannopoulos},\ and\ \citenamefont
  {Solja{\v{c}}i{\'c}}}]{wang2009observation}%
  \BibitemOpen
  \bibfield  {author} {\bibinfo {author} {\bibfnamefont {Z.}~\bibnamefont
  {Wang}}, \bibinfo {author} {\bibfnamefont {Y.}~\bibnamefont {Chong}},
  \bibinfo {author} {\bibfnamefont {J.~D.}\ \bibnamefont {Joannopoulos}},\ and\
  \bibinfo {author} {\bibfnamefont {M.}~\bibnamefont {Solja{\v{c}}i{\'c}}},\
  }\href@noop {} {\bibfield  {journal} {\bibinfo  {journal} {Nature}\ }\textbf
  {\bibinfo {volume} {461}},\ \bibinfo {pages} {772} (\bibinfo {year}
  {2009})}\BibitemShut {NoStop}%
\bibitem [{\citenamefont {Nalitov}\ \emph
  {et~al.}(2015{\natexlab{a}})\citenamefont {Nalitov}, \citenamefont
  {Solnyshkov},\ and\ \citenamefont {Malpuech}}]{nalitov2015Z}%
  \BibitemOpen
  \bibfield  {author} {\bibinfo {author} {\bibfnamefont {A.~V.}\ \bibnamefont
  {Nalitov}}, \bibinfo {author} {\bibfnamefont {D.~D.}\ \bibnamefont
  {Solnyshkov}},\ and\ \bibinfo {author} {\bibfnamefont {G.}~\bibnamefont
  {Malpuech}},\ }\href@noop {} {\bibfield  {journal} {\bibinfo  {journal}
  {Physical Review Letters}\ }\textbf {\bibinfo {volume} {114}},\ \bibinfo
  {pages} {116401} (\bibinfo {year} {2015}{\natexlab{a}})}\BibitemShut
  {NoStop}%
\bibitem [{\citenamefont {Klembt}\ \emph {et~al.}(2018)\citenamefont {Klembt},
  \citenamefont {Harder}, \citenamefont {Egorov}, \citenamefont {Winkler},
  \citenamefont {Ge}, \citenamefont {Bandres}, \citenamefont {Emmerling},
  \citenamefont {Worschech}, \citenamefont {Liew}, \citenamefont {Segev} \emph
  {et~al.}}]{klembt2018exciton}%
  \BibitemOpen
  \bibfield  {author} {\bibinfo {author} {\bibfnamefont {S.}~\bibnamefont
  {Klembt}}, \bibinfo {author} {\bibfnamefont {T.}~\bibnamefont {Harder}},
  \bibinfo {author} {\bibfnamefont {O.}~\bibnamefont {Egorov}}, \bibinfo
  {author} {\bibfnamefont {K.}~\bibnamefont {Winkler}}, \bibinfo {author}
  {\bibfnamefont {R.}~\bibnamefont {Ge}}, \bibinfo {author} {\bibfnamefont
  {M.}~\bibnamefont {Bandres}}, \bibinfo {author} {\bibfnamefont
  {M.}~\bibnamefont {Emmerling}}, \bibinfo {author} {\bibfnamefont
  {L.}~\bibnamefont {Worschech}}, \bibinfo {author} {\bibfnamefont
  {T.}~\bibnamefont {Liew}}, \bibinfo {author} {\bibfnamefont {M.}~\bibnamefont
  {Segev}}, \emph {et~al.},\ }\href@noop {} {\bibfield  {journal} {\bibinfo
  {journal} {Nature}\ }\textbf {\bibinfo {volume} {562}},\ \bibinfo {pages}
  {552} (\bibinfo {year} {2018})}\BibitemShut {NoStop}%
\bibitem [{\citenamefont {Slobozhanyuk}\ \emph {et~al.}(2017)\citenamefont
  {Slobozhanyuk}, \citenamefont {Mousavi}, \citenamefont {Ni}, \citenamefont
  {Smirnova}, \citenamefont {Kivshar},\ and\ \citenamefont
  {Khanikaev}}]{slobozhanyuk2017three}%
  \BibitemOpen
  \bibfield  {author} {\bibinfo {author} {\bibfnamefont {A.}~\bibnamefont
  {Slobozhanyuk}}, \bibinfo {author} {\bibfnamefont {S.~H.}\ \bibnamefont
  {Mousavi}}, \bibinfo {author} {\bibfnamefont {X.}~\bibnamefont {Ni}},
  \bibinfo {author} {\bibfnamefont {D.}~\bibnamefont {Smirnova}}, \bibinfo
  {author} {\bibfnamefont {Y.~S.}\ \bibnamefont {Kivshar}},\ and\ \bibinfo
  {author} {\bibfnamefont {A.~B.}\ \bibnamefont {Khanikaev}},\ }\href@noop {}
  {\bibfield  {journal} {\bibinfo  {journal} {Nature Photonics}\ }\textbf
  {\bibinfo {volume} {11}},\ \bibinfo {pages} {130} (\bibinfo {year}
  {2017})}\BibitemShut {NoStop}%
\bibitem [{\citenamefont {Solnyshkov}\ \emph {et~al.}(2016)\citenamefont
  {Solnyshkov}, \citenamefont {Nalitov},\ and\ \citenamefont
  {Malpuech}}]{Solnyshkov2016kzm}%
  \BibitemOpen
  \bibfield  {author} {\bibinfo {author} {\bibfnamefont {D.~D.}\ \bibnamefont
  {Solnyshkov}}, \bibinfo {author} {\bibfnamefont {A.~V.}\ \bibnamefont
  {Nalitov}},\ and\ \bibinfo {author} {\bibfnamefont {G.}~\bibnamefont
  {Malpuech}},\ }\href@noop {} {\bibfield  {journal} {\bibinfo  {journal}
  {Physical Review Letters}\ }\textbf {\bibinfo {volume} {116}},\ \bibinfo
  {pages} {046402} (\bibinfo {year} {2016})}\BibitemShut {NoStop}%
\bibitem [{\citenamefont {St-Jean}\ \emph {et~al.}(2017)\citenamefont
  {St-Jean}, \citenamefont {Goblot}, \citenamefont {Galopin}, \citenamefont
  {Lema{\^\i}tre}, \citenamefont {Ozawa}, \citenamefont {Le~Gratiet},
  \citenamefont {Sagnes}, \citenamefont {Bloch},\ and\ \citenamefont
  {Amo}}]{st2017lasing}%
  \BibitemOpen
  \bibfield  {author} {\bibinfo {author} {\bibfnamefont {P.}~\bibnamefont
  {St-Jean}}, \bibinfo {author} {\bibfnamefont {V.}~\bibnamefont {Goblot}},
  \bibinfo {author} {\bibfnamefont {E.}~\bibnamefont {Galopin}}, \bibinfo
  {author} {\bibfnamefont {A.}~\bibnamefont {Lema{\^\i}tre}}, \bibinfo {author}
  {\bibfnamefont {T.}~\bibnamefont {Ozawa}}, \bibinfo {author} {\bibfnamefont
  {L.}~\bibnamefont {Le~Gratiet}}, \bibinfo {author} {\bibfnamefont
  {I.}~\bibnamefont {Sagnes}}, \bibinfo {author} {\bibfnamefont
  {J.}~\bibnamefont {Bloch}},\ and\ \bibinfo {author} {\bibfnamefont
  {A.}~\bibnamefont {Amo}},\ }\href@noop {} {\bibfield  {journal} {\bibinfo
  {journal} {Nature Photonics}\ }\textbf {\bibinfo {volume} {11}},\ \bibinfo
  {pages} {651} (\bibinfo {year} {2017})}\BibitemShut {NoStop}%
\bibitem [{\citenamefont {Bahari}\ \emph {et~al.}(2017)\citenamefont {Bahari},
  \citenamefont {Ndao}, \citenamefont {Vallini}, \citenamefont {El~Amili},
  \citenamefont {Fainman},\ and\ \citenamefont
  {Kant{\'e}}}]{bahari2017nonreciprocal}%
  \BibitemOpen
  \bibfield  {author} {\bibinfo {author} {\bibfnamefont {B.}~\bibnamefont
  {Bahari}}, \bibinfo {author} {\bibfnamefont {A.}~\bibnamefont {Ndao}},
  \bibinfo {author} {\bibfnamefont {F.}~\bibnamefont {Vallini}}, \bibinfo
  {author} {\bibfnamefont {A.}~\bibnamefont {El~Amili}}, \bibinfo {author}
  {\bibfnamefont {Y.}~\bibnamefont {Fainman}},\ and\ \bibinfo {author}
  {\bibfnamefont {B.}~\bibnamefont {Kant{\'e}}},\ }\href@noop {} {\bibfield
  {journal} {\bibinfo  {journal} {Science}\ }\textbf {\bibinfo {volume}
  {358}},\ \bibinfo {pages} {636} (\bibinfo {year} {2017})}\BibitemShut
  {NoStop}%
\bibitem [{\citenamefont {Bandres}\ \emph {et~al.}(2018)\citenamefont
  {Bandres}, \citenamefont {Wittek}, \citenamefont {Harari}, \citenamefont
  {Parto}, \citenamefont {Ren}, \citenamefont {Segev}, \citenamefont
  {Christodoulides},\ and\ \citenamefont
  {Khajavikhan}}]{bandres2018topological}%
  \BibitemOpen
  \bibfield  {author} {\bibinfo {author} {\bibfnamefont {M.~A.}\ \bibnamefont
  {Bandres}}, \bibinfo {author} {\bibfnamefont {S.}~\bibnamefont {Wittek}},
  \bibinfo {author} {\bibfnamefont {G.}~\bibnamefont {Harari}}, \bibinfo
  {author} {\bibfnamefont {M.}~\bibnamefont {Parto}}, \bibinfo {author}
  {\bibfnamefont {J.}~\bibnamefont {Ren}}, \bibinfo {author} {\bibfnamefont
  {M.}~\bibnamefont {Segev}}, \bibinfo {author} {\bibfnamefont {D.~N.}\
  \bibnamefont {Christodoulides}},\ and\ \bibinfo {author} {\bibfnamefont
  {M.}~\bibnamefont {Khajavikhan}},\ }\href@noop {} {\bibfield  {journal}
  {\bibinfo  {journal} {Science}\ }\textbf {\bibinfo {volume} {359}},\ \bibinfo
  {pages} {1231} (\bibinfo {year} {2018})}\BibitemShut {NoStop}%
\bibitem [{\citenamefont {Choi}\ \emph {et~al.}(2021)\citenamefont {Choi},
  \citenamefont {Hayenga}, \citenamefont {Liu}, \citenamefont {Parto},
  \citenamefont {Bahari}, \citenamefont {Christodoulides},\ and\ \citenamefont
  {Khajavikhan}}]{choi2021room}%
  \BibitemOpen
  \bibfield  {author} {\bibinfo {author} {\bibfnamefont {J.-H.}\ \bibnamefont
  {Choi}}, \bibinfo {author} {\bibfnamefont {W.~E.}\ \bibnamefont {Hayenga}},
  \bibinfo {author} {\bibfnamefont {Y.~G.}\ \bibnamefont {Liu}}, \bibinfo
  {author} {\bibfnamefont {M.}~\bibnamefont {Parto}}, \bibinfo {author}
  {\bibfnamefont {B.}~\bibnamefont {Bahari}}, \bibinfo {author} {\bibfnamefont
  {D.~N.}\ \bibnamefont {Christodoulides}},\ and\ \bibinfo {author}
  {\bibfnamefont {M.}~\bibnamefont {Khajavikhan}},\ }\href@noop {} {\bibfield
  {journal} {\bibinfo  {journal} {Nature Communications}\ }\textbf {\bibinfo
  {volume} {12}},\ \bibinfo {pages} {3434} (\bibinfo {year}
  {2021})}\BibitemShut {NoStop}%
\bibitem [{\citenamefont {Kudyshev}\ \emph {et~al.}(2019)\citenamefont
  {Kudyshev}, \citenamefont {Kildishev}, \citenamefont {Boltasseva},\ and\
  \citenamefont {Shalaev}}]{kudyshev2019tuning}%
  \BibitemOpen
  \bibfield  {author} {\bibinfo {author} {\bibfnamefont {Z.~A.}\ \bibnamefont
  {Kudyshev}}, \bibinfo {author} {\bibfnamefont {A.~V.}\ \bibnamefont
  {Kildishev}}, \bibinfo {author} {\bibfnamefont {A.}~\bibnamefont
  {Boltasseva}},\ and\ \bibinfo {author} {\bibfnamefont {V.~M.}\ \bibnamefont
  {Shalaev}},\ }\href@noop {} {\bibfield  {journal} {\bibinfo  {journal} {ACS
  photonics}\ }\textbf {\bibinfo {volume} {6}},\ \bibinfo {pages} {1922}
  (\bibinfo {year} {2019})}\BibitemShut {NoStop}%
\bibitem [{\citenamefont {Lumer}\ \emph {et~al.}(2019)\citenamefont {Lumer},
  \citenamefont {Bandres}, \citenamefont {Heinrich}, \citenamefont {Maczewsky},
  \citenamefont {Herzig-Sheinfux}, \citenamefont {Szameit},\ and\ \citenamefont
  {Segev}}]{lumer2019light}%
  \BibitemOpen
  \bibfield  {author} {\bibinfo {author} {\bibfnamefont {Y.}~\bibnamefont
  {Lumer}}, \bibinfo {author} {\bibfnamefont {M.~A.}\ \bibnamefont {Bandres}},
  \bibinfo {author} {\bibfnamefont {M.}~\bibnamefont {Heinrich}}, \bibinfo
  {author} {\bibfnamefont {L.~J.}\ \bibnamefont {Maczewsky}}, \bibinfo {author}
  {\bibfnamefont {H.}~\bibnamefont {Herzig-Sheinfux}}, \bibinfo {author}
  {\bibfnamefont {A.}~\bibnamefont {Szameit}},\ and\ \bibinfo {author}
  {\bibfnamefont {M.}~\bibnamefont {Segev}},\ }\href@noop {} {\bibfield
  {journal} {\bibinfo  {journal} {Nature Photonics}\ }\textbf {\bibinfo
  {volume} {13}},\ \bibinfo {pages} {339} (\bibinfo {year} {2019})}\BibitemShut
  {NoStop}%
\bibitem [{\citenamefont {Chalabi}\ \emph {et~al.}(2020)\citenamefont
  {Chalabi}, \citenamefont {Barik}, \citenamefont {Mittal}, \citenamefont
  {Murphy}, \citenamefont {Hafezi},\ and\ \citenamefont
  {Waks}}]{chalabi2020guiding}%
  \BibitemOpen
  \bibfield  {author} {\bibinfo {author} {\bibfnamefont {H.}~\bibnamefont
  {Chalabi}}, \bibinfo {author} {\bibfnamefont {S.}~\bibnamefont {Barik}},
  \bibinfo {author} {\bibfnamefont {S.}~\bibnamefont {Mittal}}, \bibinfo
  {author} {\bibfnamefont {T.~E.}\ \bibnamefont {Murphy}}, \bibinfo {author}
  {\bibfnamefont {M.}~\bibnamefont {Hafezi}},\ and\ \bibinfo {author}
  {\bibfnamefont {E.}~\bibnamefont {Waks}},\ }\href@noop {} {\bibfield
  {journal} {\bibinfo  {journal} {Optica}\ }\textbf {\bibinfo {volume} {7}},\
  \bibinfo {pages} {506} (\bibinfo {year} {2020})}\BibitemShut {NoStop}%
\bibitem [{\citenamefont {Vakulenko}\ \emph {et~al.}(2023)\citenamefont
  {Vakulenko}, \citenamefont {Kiriushechkina}, \citenamefont {Smirnova},
  \citenamefont {Guddala}, \citenamefont {Komissarenko}, \citenamefont
  {Al{\`u}}, \citenamefont {Allen}, \citenamefont {Allen},\ and\ \citenamefont
  {Khanikaev}}]{vakulenko2023adiabatic}%
  \BibitemOpen
  \bibfield  {author} {\bibinfo {author} {\bibfnamefont {A.}~\bibnamefont
  {Vakulenko}}, \bibinfo {author} {\bibfnamefont {S.}~\bibnamefont
  {Kiriushechkina}}, \bibinfo {author} {\bibfnamefont {D.}~\bibnamefont
  {Smirnova}}, \bibinfo {author} {\bibfnamefont {S.}~\bibnamefont {Guddala}},
  \bibinfo {author} {\bibfnamefont {F.}~\bibnamefont {Komissarenko}}, \bibinfo
  {author} {\bibfnamefont {A.}~\bibnamefont {Al{\`u}}}, \bibinfo {author}
  {\bibfnamefont {M.}~\bibnamefont {Allen}}, \bibinfo {author} {\bibfnamefont
  {J.}~\bibnamefont {Allen}},\ and\ \bibinfo {author} {\bibfnamefont {A.~B.}\
  \bibnamefont {Khanikaev}},\ }\href@noop {} {\bibfield  {journal} {\bibinfo
  {journal} {Nature Communications}\ }\textbf {\bibinfo {volume} {14}},\
  \bibinfo {pages} {4629} (\bibinfo {year} {2023})}\BibitemShut {NoStop}%
\bibitem [{\citenamefont {Ter{\c{c}}as}\ \emph {et~al.}(2014)\citenamefont
  {Ter{\c{c}}as}, \citenamefont {Flayac}, \citenamefont {Solnyshkov},\ and\
  \citenamefont {Malpuech}}]{terccas2014non}%
  \BibitemOpen
  \bibfield  {author} {\bibinfo {author} {\bibfnamefont {H.}~\bibnamefont
  {Ter{\c{c}}as}}, \bibinfo {author} {\bibfnamefont {H.}~\bibnamefont
  {Flayac}}, \bibinfo {author} {\bibfnamefont {D.~D.}\ \bibnamefont
  {Solnyshkov}},\ and\ \bibinfo {author} {\bibfnamefont {G.}~\bibnamefont
  {Malpuech}},\ }\href@noop {} {\bibfield  {journal} {\bibinfo  {journal}
  {Physical Review Letters}\ }\textbf {\bibinfo {volume} {112}},\ \bibinfo
  {pages} {066402} (\bibinfo {year} {2014})}\BibitemShut {NoStop}%
\bibitem [{\citenamefont {Chen}\ \emph {et~al.}(2019)\citenamefont {Chen},
  \citenamefont {Zhang}, \citenamefont {Xiong}, \citenamefont {Hang},
  \citenamefont {Li}, \citenamefont {Shen},\ and\ \citenamefont
  {Chan}}]{chen2019non}%
  \BibitemOpen
  \bibfield  {author} {\bibinfo {author} {\bibfnamefont {Y.}~\bibnamefont
  {Chen}}, \bibinfo {author} {\bibfnamefont {R.-Y.}\ \bibnamefont {Zhang}},
  \bibinfo {author} {\bibfnamefont {Z.}~\bibnamefont {Xiong}}, \bibinfo
  {author} {\bibfnamefont {Z.~H.}\ \bibnamefont {Hang}}, \bibinfo {author}
  {\bibfnamefont {J.}~\bibnamefont {Li}}, \bibinfo {author} {\bibfnamefont
  {J.~Q.}\ \bibnamefont {Shen}},\ and\ \bibinfo {author} {\bibfnamefont
  {C.~T.}\ \bibnamefont {Chan}},\ }\href@noop {} {\bibfield  {journal}
  {\bibinfo  {journal} {Nature Communications}\ }\textbf {\bibinfo {volume}
  {10}},\ \bibinfo {pages} {3125} (\bibinfo {year} {2019})}\BibitemShut
  {NoStop}%
\bibitem [{\citenamefont {Polimeno}\ \emph
  {et~al.}(2021{\natexlab{a}})\citenamefont {Polimeno}, \citenamefont
  {Fieramosca}, \citenamefont {Lerario}, \citenamefont {De~Marco},
  \citenamefont {De~Giorgi}, \citenamefont {Ballarini}, \citenamefont
  {Dominici}, \citenamefont {Ardizzone}, \citenamefont {Pugliese},
  \citenamefont {Prontera} \emph {et~al.}}]{polimeno2021experimental}%
  \BibitemOpen
  \bibfield  {author} {\bibinfo {author} {\bibfnamefont {L.}~\bibnamefont
  {Polimeno}}, \bibinfo {author} {\bibfnamefont {A.}~\bibnamefont
  {Fieramosca}}, \bibinfo {author} {\bibfnamefont {G.}~\bibnamefont {Lerario}},
  \bibinfo {author} {\bibfnamefont {L.}~\bibnamefont {De~Marco}}, \bibinfo
  {author} {\bibfnamefont {M.}~\bibnamefont {De~Giorgi}}, \bibinfo {author}
  {\bibfnamefont {D.}~\bibnamefont {Ballarini}}, \bibinfo {author}
  {\bibfnamefont {L.}~\bibnamefont {Dominici}}, \bibinfo {author}
  {\bibfnamefont {V.}~\bibnamefont {Ardizzone}}, \bibinfo {author}
  {\bibfnamefont {M.}~\bibnamefont {Pugliese}}, \bibinfo {author}
  {\bibfnamefont {C.}~\bibnamefont {Prontera}}, \emph {et~al.},\ }\href@noop {}
  {\bibfield  {journal} {\bibinfo  {journal} {Optica}\ }\textbf {\bibinfo
  {volume} {8}},\ \bibinfo {pages} {1442} (\bibinfo {year}
  {2021}{\natexlab{a}})}\BibitemShut {NoStop}%
\bibitem [{\citenamefont {Yan}\ \emph {et~al.}(2023)\citenamefont {Yan},
  \citenamefont {Wang}, \citenamefont {Wang}, \citenamefont {Ma}, \citenamefont
  {Lu}, \citenamefont {Ma}, \citenamefont {Hu},\ and\ \citenamefont
  {Gong}}]{yan2023non}%
  \BibitemOpen
  \bibfield  {author} {\bibinfo {author} {\bibfnamefont {Q.}~\bibnamefont
  {Yan}}, \bibinfo {author} {\bibfnamefont {Z.}~\bibnamefont {Wang}}, \bibinfo
  {author} {\bibfnamefont {D.}~\bibnamefont {Wang}}, \bibinfo {author}
  {\bibfnamefont {R.}~\bibnamefont {Ma}}, \bibinfo {author} {\bibfnamefont
  {C.}~\bibnamefont {Lu}}, \bibinfo {author} {\bibfnamefont {G.}~\bibnamefont
  {Ma}}, \bibinfo {author} {\bibfnamefont {X.}~\bibnamefont {Hu}},\ and\
  \bibinfo {author} {\bibfnamefont {Q.}~\bibnamefont {Gong}},\ }\href@noop {}
  {\bibfield  {journal} {\bibinfo  {journal} {Advances in Optics and
  Photonics}\ }\textbf {\bibinfo {volume} {15}},\ \bibinfo {pages} {907}
  (\bibinfo {year} {2023})}\BibitemShut {NoStop}%
\bibitem [{\citenamefont {Yang}\ \emph {et~al.}(2024)\citenamefont {Yang},
  \citenamefont {Yang}, \citenamefont {Ma}, \citenamefont {Li}, \citenamefont
  {Zhang},\ and\ \citenamefont {Chan}}]{yang2024non}%
  \BibitemOpen
  \bibfield  {author} {\bibinfo {author} {\bibfnamefont {Y.}~\bibnamefont
  {Yang}}, \bibinfo {author} {\bibfnamefont {B.}~\bibnamefont {Yang}}, \bibinfo
  {author} {\bibfnamefont {G.}~\bibnamefont {Ma}}, \bibinfo {author}
  {\bibfnamefont {J.}~\bibnamefont {Li}}, \bibinfo {author} {\bibfnamefont
  {S.}~\bibnamefont {Zhang}},\ and\ \bibinfo {author} {\bibfnamefont
  {C.}~\bibnamefont {Chan}},\ }\href@noop {} {\bibfield  {journal} {\bibinfo
  {journal} {Science}\ }\textbf {\bibinfo {volume} {383}},\ \bibinfo {pages}
  {844} (\bibinfo {year} {2024})}\BibitemShut {NoStop}%
\bibitem [{\citenamefont {Yang}\ \emph
  {et~al.}(2020{\natexlab{a}})\citenamefont {Yang}, \citenamefont {Zhen},
  \citenamefont {Joannopoulos},\ and\ \citenamefont
  {Solja{\v{c}}i{\'c}}}]{yang2020non}%
  \BibitemOpen
  \bibfield  {author} {\bibinfo {author} {\bibfnamefont {Y.}~\bibnamefont
  {Yang}}, \bibinfo {author} {\bibfnamefont {B.}~\bibnamefont {Zhen}}, \bibinfo
  {author} {\bibfnamefont {J.~D.}\ \bibnamefont {Joannopoulos}},\ and\ \bibinfo
  {author} {\bibfnamefont {M.}~\bibnamefont {Solja{\v{c}}i{\'c}}},\ }\href@noop
  {} {\bibfield  {journal} {\bibinfo  {journal} {Light: Science \&
  Applications}\ }\textbf {\bibinfo {volume} {9}},\ \bibinfo {pages} {177}
  (\bibinfo {year} {2020}{\natexlab{a}})}\BibitemShut {NoStop}%
\bibitem [{\citenamefont {Ni}\ \emph {et~al.}(2018)\citenamefont {Ni},
  \citenamefont {Purtseladze}, \citenamefont {Smirnova}, \citenamefont
  {Slobozhanyuk}, \citenamefont {Alù},\ and\ \citenamefont
  {Khanikaev}}]{ni2018spin}%
  \BibitemOpen
  \bibfield  {author} {\bibinfo {author} {\bibfnamefont {X.}~\bibnamefont
  {Ni}}, \bibinfo {author} {\bibfnamefont {D.}~\bibnamefont {Purtseladze}},
  \bibinfo {author} {\bibfnamefont {D.~A.}\ \bibnamefont {Smirnova}}, \bibinfo
  {author} {\bibfnamefont {A.}~\bibnamefont {Slobozhanyuk}}, \bibinfo {author}
  {\bibfnamefont {A.}~\bibnamefont {Alù}},\ and\ \bibinfo {author}
  {\bibfnamefont {A.~B.}\ \bibnamefont {Khanikaev}},\ }\href@noop {} {\bibfield
   {journal} {\bibinfo  {journal} {Science Advances}\ }\textbf {\bibinfo
  {volume} {4}} (\bibinfo {year} {2018})}\BibitemShut {NoStop}%
\bibitem [{\citenamefont {Jia}\ \emph {et~al.}(2019)\citenamefont {Jia},
  \citenamefont {Zhang}, \citenamefont {Gao}, \citenamefont {Guo},
  \citenamefont {Yang}, \citenamefont {Hu}, \citenamefont {Bi}, \citenamefont
  {Xiang}, \citenamefont {Liu},\ and\ \citenamefont
  {Zhang}}]{jia2019observation}%
  \BibitemOpen
  \bibfield  {author} {\bibinfo {author} {\bibfnamefont {H.}~\bibnamefont
  {Jia}}, \bibinfo {author} {\bibfnamefont {R.}~\bibnamefont {Zhang}}, \bibinfo
  {author} {\bibfnamefont {W.}~\bibnamefont {Gao}}, \bibinfo {author}
  {\bibfnamefont {Q.}~\bibnamefont {Guo}}, \bibinfo {author} {\bibfnamefont
  {B.}~\bibnamefont {Yang}}, \bibinfo {author} {\bibfnamefont {J.}~\bibnamefont
  {Hu}}, \bibinfo {author} {\bibfnamefont {Y.}~\bibnamefont {Bi}}, \bibinfo
  {author} {\bibfnamefont {Y.}~\bibnamefont {Xiang}}, \bibinfo {author}
  {\bibfnamefont {C.}~\bibnamefont {Liu}},\ and\ \bibinfo {author}
  {\bibfnamefont {S.}~\bibnamefont {Zhang}},\ }\href@noop {} {\bibfield
  {journal} {\bibinfo  {journal} {Science}\ }\textbf {\bibinfo {volume}
  {363}},\ \bibinfo {pages} {148} (\bibinfo {year} {2019})}\BibitemShut
  {NoStop}%
\bibitem [{\citenamefont {Aidelsburger}\ \emph {et~al.}(2018)\citenamefont
  {Aidelsburger}, \citenamefont {Nascimbene},\ and\ \citenamefont
  {Goldman}}]{aidelsburger2018artificial}%
  \BibitemOpen
  \bibfield  {author} {\bibinfo {author} {\bibfnamefont {M.}~\bibnamefont
  {Aidelsburger}}, \bibinfo {author} {\bibfnamefont {S.}~\bibnamefont
  {Nascimbene}},\ and\ \bibinfo {author} {\bibfnamefont {N.}~\bibnamefont
  {Goldman}},\ }\href@noop {} {\bibfield  {journal} {\bibinfo  {journal}
  {Comptes Rendus Physique}\ }\textbf {\bibinfo {volume} {19}},\ \bibinfo
  {pages} {394} (\bibinfo {year} {2018})}\BibitemShut {NoStop}%
\bibitem [{\citenamefont {Price}\ \emph {et~al.}(2014)\citenamefont {Price},
  \citenamefont {Ozawa},\ and\ \citenamefont {Carusotto}}]{Price2014}%
  \BibitemOpen
  \bibfield  {author} {\bibinfo {author} {\bibfnamefont {H.~M.}\ \bibnamefont
  {Price}}, \bibinfo {author} {\bibfnamefont {T.}~\bibnamefont {Ozawa}},\ and\
  \bibinfo {author} {\bibfnamefont {I.}~\bibnamefont {Carusotto}},\ }\href@noop
  {} {\bibfield  {journal} {\bibinfo  {journal} {Physical Review Letters}\
  }\textbf {\bibinfo {volume} {113}},\ \bibinfo {pages} {190403} (\bibinfo
  {year} {2014})}\BibitemShut {NoStop}%
\bibitem [{\citenamefont {Rechtsman}\ \emph
  {et~al.}(2013{\natexlab{a}})\citenamefont {Rechtsman}, \citenamefont
  {Zeuner}, \citenamefont {Plotnik}, \citenamefont {Lumer}, \citenamefont
  {Podolsky}, \citenamefont {Dreisow}, \citenamefont {Nolte}, \citenamefont
  {Segev},\ and\ \citenamefont {Szameit}}]{rechtsman2013photonic}%
  \BibitemOpen
  \bibfield  {author} {\bibinfo {author} {\bibfnamefont {M.~C.}\ \bibnamefont
  {Rechtsman}}, \bibinfo {author} {\bibfnamefont {J.~M.}\ \bibnamefont
  {Zeuner}}, \bibinfo {author} {\bibfnamefont {Y.}~\bibnamefont {Plotnik}},
  \bibinfo {author} {\bibfnamefont {Y.}~\bibnamefont {Lumer}}, \bibinfo
  {author} {\bibfnamefont {D.}~\bibnamefont {Podolsky}}, \bibinfo {author}
  {\bibfnamefont {F.}~\bibnamefont {Dreisow}}, \bibinfo {author} {\bibfnamefont
  {S.}~\bibnamefont {Nolte}}, \bibinfo {author} {\bibfnamefont
  {M.}~\bibnamefont {Segev}},\ and\ \bibinfo {author} {\bibfnamefont
  {A.}~\bibnamefont {Szameit}},\ }\href@noop {} {\bibfield  {journal} {\bibinfo
   {journal} {Nature}\ }\textbf {\bibinfo {volume} {496}},\ \bibinfo {pages}
  {196} (\bibinfo {year} {2013}{\natexlab{a}})}\BibitemShut {NoStop}%
\bibitem [{\citenamefont {Hey}\ and\ \citenamefont
  {Li}(2018)}]{hey2018advances}%
  \BibitemOpen
  \bibfield  {author} {\bibinfo {author} {\bibfnamefont {D.}~\bibnamefont
  {Hey}}\ and\ \bibinfo {author} {\bibfnamefont {E.}~\bibnamefont {Li}},\
  }\href@noop {} {\bibfield  {journal} {\bibinfo  {journal} {Royal Society Open
  Science}\ }\textbf {\bibinfo {volume} {5}},\ \bibinfo {pages} {172447}
  (\bibinfo {year} {2018})}\BibitemShut {NoStop}%
\bibitem [{\citenamefont {Yang}\ \emph
  {et~al.}(2020{\natexlab{b}})\citenamefont {Yang}, \citenamefont {Lustig},
  \citenamefont {Lumer},\ and\ \citenamefont {Segev}}]{yang2020photonic}%
  \BibitemOpen
  \bibfield  {author} {\bibinfo {author} {\bibfnamefont {Z.}~\bibnamefont
  {Yang}}, \bibinfo {author} {\bibfnamefont {E.}~\bibnamefont {Lustig}},
  \bibinfo {author} {\bibfnamefont {Y.}~\bibnamefont {Lumer}},\ and\ \bibinfo
  {author} {\bibfnamefont {M.}~\bibnamefont {Segev}},\ }\href@noop {}
  {\bibfield  {journal} {\bibinfo  {journal} {Light: Science \& Applications}\
  }\textbf {\bibinfo {volume} {9}},\ \bibinfo {pages} {128} (\bibinfo {year}
  {2020}{\natexlab{b}})}\BibitemShut {NoStop}%
\bibitem [{\citenamefont {Ozawa}\ \emph {et~al.}(2016)\citenamefont {Ozawa},
  \citenamefont {Price}, \citenamefont {Goldman}, \citenamefont {Zilberberg},\
  and\ \citenamefont {Carusotto}}]{ozawa2016synthetic}%
  \BibitemOpen
  \bibfield  {author} {\bibinfo {author} {\bibfnamefont {T.}~\bibnamefont
  {Ozawa}}, \bibinfo {author} {\bibfnamefont {H.~M.}\ \bibnamefont {Price}},
  \bibinfo {author} {\bibfnamefont {N.}~\bibnamefont {Goldman}}, \bibinfo
  {author} {\bibfnamefont {O.}~\bibnamefont {Zilberberg}},\ and\ \bibinfo
  {author} {\bibfnamefont {I.}~\bibnamefont {Carusotto}},\ }\href@noop {}
  {\bibfield  {journal} {\bibinfo  {journal} {Physical Review A}\ }\textbf
  {\bibinfo {volume} {93}},\ \bibinfo {pages} {043827} (\bibinfo {year}
  {2016})}\BibitemShut {NoStop}%
\bibitem [{\citenamefont {Lustig}\ \emph {et~al.}(2019)\citenamefont {Lustig},
  \citenamefont {Weimann}, \citenamefont {Plotnik}, \citenamefont {Lumer},
  \citenamefont {Bandres}, \citenamefont {Szameit},\ and\ \citenamefont
  {Segev}}]{lustig2019photonic}%
  \BibitemOpen
  \bibfield  {author} {\bibinfo {author} {\bibfnamefont {E.}~\bibnamefont
  {Lustig}}, \bibinfo {author} {\bibfnamefont {S.}~\bibnamefont {Weimann}},
  \bibinfo {author} {\bibfnamefont {Y.}~\bibnamefont {Plotnik}}, \bibinfo
  {author} {\bibfnamefont {Y.}~\bibnamefont {Lumer}}, \bibinfo {author}
  {\bibfnamefont {M.~A.}\ \bibnamefont {Bandres}}, \bibinfo {author}
  {\bibfnamefont {A.}~\bibnamefont {Szameit}},\ and\ \bibinfo {author}
  {\bibfnamefont {M.}~\bibnamefont {Segev}},\ }\href@noop {} {\bibfield
  {journal} {\bibinfo  {journal} {Nature}\ }\textbf {\bibinfo {volume} {567}},\
  \bibinfo {pages} {356} (\bibinfo {year} {2019})}\BibitemShut {NoStop}%
\bibitem [{\citenamefont {Dutt}\ \emph {et~al.}(2020)\citenamefont {Dutt},
  \citenamefont {Lin}, \citenamefont {Yuan}, \citenamefont {Minkov},
  \citenamefont {Xiao},\ and\ \citenamefont {Fan}}]{dutt2020single}%
  \BibitemOpen
  \bibfield  {author} {\bibinfo {author} {\bibfnamefont {A.}~\bibnamefont
  {Dutt}}, \bibinfo {author} {\bibfnamefont {Q.}~\bibnamefont {Lin}}, \bibinfo
  {author} {\bibfnamefont {L.}~\bibnamefont {Yuan}}, \bibinfo {author}
  {\bibfnamefont {M.}~\bibnamefont {Minkov}}, \bibinfo {author} {\bibfnamefont
  {M.}~\bibnamefont {Xiao}},\ and\ \bibinfo {author} {\bibfnamefont
  {S.}~\bibnamefont {Fan}},\ }\href@noop {} {\bibfield  {journal} {\bibinfo
  {journal} {Science}\ }\textbf {\bibinfo {volume} {367}},\ \bibinfo {pages}
  {59} (\bibinfo {year} {2020})}\BibitemShut {NoStop}%
\bibitem [{\citenamefont {Nalitov}\ \emph
  {et~al.}(2015{\natexlab{b}})\citenamefont {Nalitov}, \citenamefont
  {Malpuech}, \citenamefont {Ter{\c{c}}as},\ and\ \citenamefont
  {Solnyshkov}}]{nalitov2015spin}%
  \BibitemOpen
  \bibfield  {author} {\bibinfo {author} {\bibfnamefont {A.~V.}\ \bibnamefont
  {Nalitov}}, \bibinfo {author} {\bibfnamefont {G.}~\bibnamefont {Malpuech}},
  \bibinfo {author} {\bibfnamefont {H.}~\bibnamefont {Ter{\c{c}}as}},\ and\
  \bibinfo {author} {\bibfnamefont {D.~D.}\ \bibnamefont {Solnyshkov}},\
  }\href@noop {} {\bibfield  {journal} {\bibinfo  {journal} {Physical Review
  Letters}\ }\textbf {\bibinfo {volume} {114}},\ \bibinfo {pages} {026803}
  (\bibinfo {year} {2015}{\natexlab{b}})}\BibitemShut {NoStop}%
\bibitem [{\citenamefont {Whittaker}\ \emph {et~al.}(2021)\citenamefont
  {Whittaker}, \citenamefont {Dowling}, \citenamefont {Nalitov}, \citenamefont
  {Yulin}, \citenamefont {Royall}, \citenamefont {Clarke}, \citenamefont
  {Skolnick}, \citenamefont {Shelykh},\ and\ \citenamefont
  {Krizhanovskii}}]{whittaker2021optical}%
  \BibitemOpen
  \bibfield  {author} {\bibinfo {author} {\bibfnamefont {C.}~\bibnamefont
  {Whittaker}}, \bibinfo {author} {\bibfnamefont {T.}~\bibnamefont {Dowling}},
  \bibinfo {author} {\bibfnamefont {A.~V.}\ \bibnamefont {Nalitov}}, \bibinfo
  {author} {\bibfnamefont {A.~V.}\ \bibnamefont {Yulin}}, \bibinfo {author}
  {\bibfnamefont {B.}~\bibnamefont {Royall}}, \bibinfo {author} {\bibfnamefont
  {E.}~\bibnamefont {Clarke}}, \bibinfo {author} {\bibfnamefont {M.~S.}\
  \bibnamefont {Skolnick}}, \bibinfo {author} {\bibfnamefont {I.~A.}\
  \bibnamefont {Shelykh}},\ and\ \bibinfo {author} {\bibfnamefont {D.~N.}\
  \bibnamefont {Krizhanovskii}},\ }\href@noop {} {\bibfield  {journal}
  {\bibinfo  {journal} {Nature Photonics}\ }\textbf {\bibinfo {volume} {15}},\
  \bibinfo {pages} {193} (\bibinfo {year} {2021})}\BibitemShut {NoStop}%
\bibitem [{\citenamefont {Rechtsman}\ \emph
  {et~al.}(2013{\natexlab{b}})\citenamefont {Rechtsman}, \citenamefont
  {Zeuner}, \citenamefont {T{\"u}nnermann}, \citenamefont {Nolte},
  \citenamefont {Segev},\ and\ \citenamefont {Szameit}}]{rechtsman2013strain}%
  \BibitemOpen
  \bibfield  {author} {\bibinfo {author} {\bibfnamefont {M.~C.}\ \bibnamefont
  {Rechtsman}}, \bibinfo {author} {\bibfnamefont {J.~M.}\ \bibnamefont
  {Zeuner}}, \bibinfo {author} {\bibfnamefont {A.}~\bibnamefont
  {T{\"u}nnermann}}, \bibinfo {author} {\bibfnamefont {S.}~\bibnamefont
  {Nolte}}, \bibinfo {author} {\bibfnamefont {M.}~\bibnamefont {Segev}},\ and\
  \bibinfo {author} {\bibfnamefont {A.}~\bibnamefont {Szameit}},\ }\href@noop
  {} {\bibfield  {journal} {\bibinfo  {journal} {Nature Photonics}\ }\textbf
  {\bibinfo {volume} {7}},\ \bibinfo {pages} {153} (\bibinfo {year}
  {2013}{\natexlab{b}})}\BibitemShut {NoStop}%
\bibitem [{\citenamefont {Jamadi}\ \emph {et~al.}(2020)\citenamefont {Jamadi},
  \citenamefont {Rozas}, \citenamefont {Salerno}, \citenamefont
  {Mili{\'c}evi{\'c}}, \citenamefont {Ozawa}, \citenamefont {Sagnes},
  \citenamefont {Lema{\^\i}tre}, \citenamefont {Le~Gratiet}, \citenamefont
  {Harouri}, \citenamefont {Carusotto} \emph {et~al.}}]{jamadi2020direct}%
  \BibitemOpen
  \bibfield  {author} {\bibinfo {author} {\bibfnamefont {O.}~\bibnamefont
  {Jamadi}}, \bibinfo {author} {\bibfnamefont {E.}~\bibnamefont {Rozas}},
  \bibinfo {author} {\bibfnamefont {G.}~\bibnamefont {Salerno}}, \bibinfo
  {author} {\bibfnamefont {M.}~\bibnamefont {Mili{\'c}evi{\'c}}}, \bibinfo
  {author} {\bibfnamefont {T.}~\bibnamefont {Ozawa}}, \bibinfo {author}
  {\bibfnamefont {I.}~\bibnamefont {Sagnes}}, \bibinfo {author} {\bibfnamefont
  {A.}~\bibnamefont {Lema{\^\i}tre}}, \bibinfo {author} {\bibfnamefont
  {L.}~\bibnamefont {Le~Gratiet}}, \bibinfo {author} {\bibfnamefont
  {A.}~\bibnamefont {Harouri}}, \bibinfo {author} {\bibfnamefont
  {I.}~\bibnamefont {Carusotto}}, \emph {et~al.},\ }\href@noop {} {\bibfield
  {journal} {\bibinfo  {journal} {Light: Science \& Applications}\ }\textbf
  {\bibinfo {volume} {9}},\ \bibinfo {pages} {144} (\bibinfo {year}
  {2020})}\BibitemShut {NoStop}%
\bibitem [{\citenamefont {Barczyk}\ \emph {et~al.}(2024)\citenamefont
  {Barczyk}, \citenamefont {Kuipers},\ and\ \citenamefont
  {Verhagen}}]{barczyk2024observation}%
  \BibitemOpen
  \bibfield  {author} {\bibinfo {author} {\bibfnamefont {R.}~\bibnamefont
  {Barczyk}}, \bibinfo {author} {\bibfnamefont {L.}~\bibnamefont {Kuipers}},\
  and\ \bibinfo {author} {\bibfnamefont {E.}~\bibnamefont {Verhagen}},\
  }\href@noop {} {\bibfield  {journal} {\bibinfo  {journal} {Nature Photonics}\
  }\textbf {\bibinfo {volume} {18}},\ \bibinfo {pages} {574–579} (\bibinfo
  {year} {2024})}\BibitemShut {NoStop}%
\bibitem [{\citenamefont {Hafezi}\ \emph {et~al.}(2011)\citenamefont {Hafezi},
  \citenamefont {Demler}, \citenamefont {Lukin},\ and\ \citenamefont
  {Taylor}}]{hafezi2011robust}%
  \BibitemOpen
  \bibfield  {author} {\bibinfo {author} {\bibfnamefont {M.}~\bibnamefont
  {Hafezi}}, \bibinfo {author} {\bibfnamefont {E.~A.}\ \bibnamefont {Demler}},
  \bibinfo {author} {\bibfnamefont {M.~D.}\ \bibnamefont {Lukin}},\ and\
  \bibinfo {author} {\bibfnamefont {J.~M.}\ \bibnamefont {Taylor}},\
  }\href@noop {} {\bibfield  {journal} {\bibinfo  {journal} {Nature Physics}\
  }\textbf {\bibinfo {volume} {7}},\ \bibinfo {pages} {907} (\bibinfo {year}
  {2011})}\BibitemShut {NoStop}%
\bibitem [{\citenamefont {Umucal{\i}lar}\ and\ \citenamefont
  {Carusotto}(2011)}]{umucalilar2011artificial}%
  \BibitemOpen
  \bibfield  {author} {\bibinfo {author} {\bibfnamefont {R.~O.}\ \bibnamefont
  {Umucal{\i}lar}}\ and\ \bibinfo {author} {\bibfnamefont {I.}~\bibnamefont
  {Carusotto}},\ }\href@noop {} {\bibfield  {journal} {\bibinfo  {journal}
  {Physical Review A}\ }\textbf {\bibinfo {volume} {84}},\ \bibinfo {pages}
  {043804} (\bibinfo {year} {2011})}\BibitemShut {NoStop}%
\bibitem [{\citenamefont {Harari}\ \emph {et~al.}(2018)\citenamefont {Harari},
  \citenamefont {Bandres}, \citenamefont {Lumer}, \citenamefont {Rechtsman},
  \citenamefont {Chong}, \citenamefont {Khajavikhan}, \citenamefont
  {Christodoulides},\ and\ \citenamefont {Segev}}]{harari2018topological}%
  \BibitemOpen
  \bibfield  {author} {\bibinfo {author} {\bibfnamefont {G.}~\bibnamefont
  {Harari}}, \bibinfo {author} {\bibfnamefont {M.~A.}\ \bibnamefont {Bandres}},
  \bibinfo {author} {\bibfnamefont {Y.}~\bibnamefont {Lumer}}, \bibinfo
  {author} {\bibfnamefont {M.~C.}\ \bibnamefont {Rechtsman}}, \bibinfo {author}
  {\bibfnamefont {Y.~D.}\ \bibnamefont {Chong}}, \bibinfo {author}
  {\bibfnamefont {M.}~\bibnamefont {Khajavikhan}}, \bibinfo {author}
  {\bibfnamefont {D.~N.}\ \bibnamefont {Christodoulides}},\ and\ \bibinfo
  {author} {\bibfnamefont {M.}~\bibnamefont {Segev}},\ }\href@noop {}
  {\bibfield  {journal} {\bibinfo  {journal} {Science}\ }\textbf {\bibinfo
  {volume} {359}},\ \bibinfo {pages} {1230} (\bibinfo {year}
  {2018})}\BibitemShut {NoStop}%
\bibitem [{\citenamefont {Rechci{\'n}ska}\ \emph {et~al.}(2019)\citenamefont
  {Rechci{\'n}ska}, \citenamefont {Kr{\'o}l}, \citenamefont {Mazur},
  \citenamefont {Morawiak}, \citenamefont {Mirek}, \citenamefont {{\L}empicka},
  \citenamefont {Bardyszewski}, \citenamefont {Matuszewski}, \citenamefont
  {Kula}, \citenamefont {Piecek} \emph {et~al.}}]{rechcinska2019engineering}%
  \BibitemOpen
  \bibfield  {author} {\bibinfo {author} {\bibfnamefont {K.}~\bibnamefont
  {Rechci{\'n}ska}}, \bibinfo {author} {\bibfnamefont {M.}~\bibnamefont
  {Kr{\'o}l}}, \bibinfo {author} {\bibfnamefont {R.}~\bibnamefont {Mazur}},
  \bibinfo {author} {\bibfnamefont {P.}~\bibnamefont {Morawiak}}, \bibinfo
  {author} {\bibfnamefont {R.}~\bibnamefont {Mirek}}, \bibinfo {author}
  {\bibfnamefont {K.}~\bibnamefont {{\L}empicka}}, \bibinfo {author}
  {\bibfnamefont {W.}~\bibnamefont {Bardyszewski}}, \bibinfo {author}
  {\bibfnamefont {M.}~\bibnamefont {Matuszewski}}, \bibinfo {author}
  {\bibfnamefont {P.}~\bibnamefont {Kula}}, \bibinfo {author} {\bibfnamefont
  {W.}~\bibnamefont {Piecek}}, \emph {et~al.},\ }\href@noop {} {\bibfield
  {journal} {\bibinfo  {journal} {Science}\ }\textbf {\bibinfo {volume}
  {366}},\ \bibinfo {pages} {727} (\bibinfo {year} {2019})}\BibitemShut
  {NoStop}%
\bibitem [{\citenamefont {Ren}\ \emph {et~al.}(2021)\citenamefont {Ren},
  \citenamefont {Liao}, \citenamefont {Li}, \citenamefont {Li}, \citenamefont
  {Bleu}, \citenamefont {Malpuech}, \citenamefont {Yao}, \citenamefont {Fu},\
  and\ \citenamefont {Solnyshkov}}]{ren2021nontrivial}%
  \BibitemOpen
  \bibfield  {author} {\bibinfo {author} {\bibfnamefont {J.}~\bibnamefont
  {Ren}}, \bibinfo {author} {\bibfnamefont {Q.}~\bibnamefont {Liao}}, \bibinfo
  {author} {\bibfnamefont {F.}~\bibnamefont {Li}}, \bibinfo {author}
  {\bibfnamefont {Y.}~\bibnamefont {Li}}, \bibinfo {author} {\bibfnamefont
  {O.}~\bibnamefont {Bleu}}, \bibinfo {author} {\bibfnamefont {G.}~\bibnamefont
  {Malpuech}}, \bibinfo {author} {\bibfnamefont {J.}~\bibnamefont {Yao}},
  \bibinfo {author} {\bibfnamefont {H.}~\bibnamefont {Fu}},\ and\ \bibinfo
  {author} {\bibfnamefont {D.}~\bibnamefont {Solnyshkov}},\ }\href@noop {}
  {\bibfield  {journal} {\bibinfo  {journal} {Nature Communications}\ }\textbf
  {\bibinfo {volume} {12}},\ \bibinfo {pages} {689} (\bibinfo {year}
  {2021})}\BibitemShut {NoStop}%
\bibitem [{\citenamefont {Liu}\ \emph {et~al.}(2019)\citenamefont {Liu},
  \citenamefont {Xu}, \citenamefont {Wang}, \citenamefont {Hang},\ and\
  \citenamefont {Li}}]{liu2019polarization}%
  \BibitemOpen
  \bibfield  {author} {\bibinfo {author} {\bibfnamefont {F.}~\bibnamefont
  {Liu}}, \bibinfo {author} {\bibfnamefont {T.}~\bibnamefont {Xu}}, \bibinfo
  {author} {\bibfnamefont {S.}~\bibnamefont {Wang}}, \bibinfo {author}
  {\bibfnamefont {Z.~H.}\ \bibnamefont {Hang}},\ and\ \bibinfo {author}
  {\bibfnamefont {J.}~\bibnamefont {Li}},\ }\href@noop {} {\bibfield  {journal}
  {\bibinfo  {journal} {Advanced Optical Materials}\ }\textbf {\bibinfo
  {volume} {7}},\ \bibinfo {pages} {1801582} (\bibinfo {year}
  {2019})}\BibitemShut {NoStop}%
\bibitem [{\citenamefont {Kr{\'o}l}\ \emph {et~al.}(2021)\citenamefont
  {Kr{\'o}l}, \citenamefont {Rechci{\'n}ska}, \citenamefont {Sigurdsson},
  \citenamefont {Oliwa}, \citenamefont {Mazur}, \citenamefont {Morawiak},
  \citenamefont {Piecek}, \citenamefont {Kula}, \citenamefont {Lagoudakis},
  \citenamefont {Matuszewski} \emph {et~al.}}]{krol2021realizing}%
  \BibitemOpen
  \bibfield  {author} {\bibinfo {author} {\bibfnamefont {M.}~\bibnamefont
  {Kr{\'o}l}}, \bibinfo {author} {\bibfnamefont {K.}~\bibnamefont
  {Rechci{\'n}ska}}, \bibinfo {author} {\bibfnamefont {H.}~\bibnamefont
  {Sigurdsson}}, \bibinfo {author} {\bibfnamefont {P.}~\bibnamefont {Oliwa}},
  \bibinfo {author} {\bibfnamefont {R.}~\bibnamefont {Mazur}}, \bibinfo
  {author} {\bibfnamefont {P.}~\bibnamefont {Morawiak}}, \bibinfo {author}
  {\bibfnamefont {W.}~\bibnamefont {Piecek}}, \bibinfo {author} {\bibfnamefont
  {P.}~\bibnamefont {Kula}}, \bibinfo {author} {\bibfnamefont {P.~G.}\
  \bibnamefont {Lagoudakis}}, \bibinfo {author} {\bibfnamefont
  {M.}~\bibnamefont {Matuszewski}}, \emph {et~al.},\ }\href@noop {} {\bibfield
  {journal} {\bibinfo  {journal} {Physical Review Letters}\ }\textbf {\bibinfo
  {volume} {127}},\ \bibinfo {pages} {190401} (\bibinfo {year}
  {2021})}\BibitemShut {NoStop}%
\bibitem [{\citenamefont {Muszy{\'n}ski}\ \emph {et~al.}(2022)\citenamefont
  {Muszy{\'n}ski}, \citenamefont {Kr\'ol}, \citenamefont {Rechci{\'n}ska},
  \citenamefont {Oliwa}, \citenamefont {K\k{e}dziora}, \citenamefont
  {\L{}empicka-Mirek}, \citenamefont {Mazur}, \citenamefont {Morawiak},
  \citenamefont {Piecek}, \citenamefont {Kula}, \citenamefont {Lagoudakis},
  \citenamefont {Pi\k{e}tka},\ and\ \citenamefont
  {Szczytko}}]{muszynski2022realizing}%
  \BibitemOpen
  \bibfield  {author} {\bibinfo {author} {\bibfnamefont {M.}~\bibnamefont
  {Muszy{\'n}ski}}, \bibinfo {author} {\bibfnamefont {M.}~\bibnamefont
  {Kr\'ol}}, \bibinfo {author} {\bibfnamefont {K.}~\bibnamefont
  {Rechci{\'n}ska}}, \bibinfo {author} {\bibfnamefont {P.}~\bibnamefont
  {Oliwa}}, \bibinfo {author} {\bibfnamefont {M.}~\bibnamefont {K\k{e}dziora}},
  \bibinfo {author} {\bibfnamefont {K.}~\bibnamefont {\L{}empicka-Mirek}},
  \bibinfo {author} {\bibfnamefont {R.}~\bibnamefont {Mazur}}, \bibinfo
  {author} {\bibfnamefont {P.}~\bibnamefont {Morawiak}}, \bibinfo {author}
  {\bibfnamefont {W.}~\bibnamefont {Piecek}}, \bibinfo {author} {\bibfnamefont
  {P.}~\bibnamefont {Kula}}, \bibinfo {author} {\bibfnamefont {P.~G.}\
  \bibnamefont {Lagoudakis}}, \bibinfo {author} {\bibfnamefont
  {B.}~\bibnamefont {Pi\k{e}tka}},\ and\ \bibinfo {author} {\bibfnamefont
  {J.}~\bibnamefont {Szczytko}},\ }\href@noop {} {\bibfield  {journal}
  {\bibinfo  {journal} {Physical Review Applied}\ }\textbf {\bibinfo {volume}
  {17}},\ \bibinfo {pages} {014041} (\bibinfo {year} {2022})}\BibitemShut
  {NoStop}%
\bibitem [{\citenamefont {Polimeno}\ \emph
  {et~al.}(2021{\natexlab{b}})\citenamefont {Polimeno}, \citenamefont
  {Lerario}, \citenamefont {De~Giorgi}, \citenamefont {De~Marco}, \citenamefont
  {Dominici}, \citenamefont {Todisco}, \citenamefont {Coriolano}, \citenamefont
  {Ardizzone}, \citenamefont {Pugliese}, \citenamefont {Prontera} \emph
  {et~al.}}]{polimeno2021tuning}%
  \BibitemOpen
  \bibfield  {author} {\bibinfo {author} {\bibfnamefont {L.}~\bibnamefont
  {Polimeno}}, \bibinfo {author} {\bibfnamefont {G.}~\bibnamefont {Lerario}},
  \bibinfo {author} {\bibfnamefont {M.}~\bibnamefont {De~Giorgi}}, \bibinfo
  {author} {\bibfnamefont {L.}~\bibnamefont {De~Marco}}, \bibinfo {author}
  {\bibfnamefont {L.}~\bibnamefont {Dominici}}, \bibinfo {author}
  {\bibfnamefont {F.}~\bibnamefont {Todisco}}, \bibinfo {author} {\bibfnamefont
  {A.}~\bibnamefont {Coriolano}}, \bibinfo {author} {\bibfnamefont
  {V.}~\bibnamefont {Ardizzone}}, \bibinfo {author} {\bibfnamefont
  {M.}~\bibnamefont {Pugliese}}, \bibinfo {author} {\bibfnamefont {C.~T.}\
  \bibnamefont {Prontera}}, \emph {et~al.},\ }\href@noop {} {\bibfield
  {journal} {\bibinfo  {journal} {Nature Nanotechnology}\ }\textbf {\bibinfo
  {volume} {16}},\ \bibinfo {pages} {1349} (\bibinfo {year}
  {2021}{\natexlab{b}})}\BibitemShut {NoStop}%
\bibitem [{\citenamefont {{\L}empicka-Mirek}\ \emph {et~al.}(2022)\citenamefont
  {{\L}empicka-Mirek}, \citenamefont {Kr{\'o}l}, \citenamefont {Sigurdsson},
  \citenamefont {Wincukiewicz}, \citenamefont {Morawiak}, \citenamefont
  {Mazur}, \citenamefont {Muszy{\'n}ski}, \citenamefont {Piecek}, \citenamefont
  {Kula}, \citenamefont {Stefaniuk} \emph {et~al.}}]{lempicka2022electrically}%
  \BibitemOpen
  \bibfield  {author} {\bibinfo {author} {\bibfnamefont {K.}~\bibnamefont
  {{\L}empicka-Mirek}}, \bibinfo {author} {\bibfnamefont {M.}~\bibnamefont
  {Kr{\'o}l}}, \bibinfo {author} {\bibfnamefont {H.}~\bibnamefont
  {Sigurdsson}}, \bibinfo {author} {\bibfnamefont {A.}~\bibnamefont
  {Wincukiewicz}}, \bibinfo {author} {\bibfnamefont {P.}~\bibnamefont
  {Morawiak}}, \bibinfo {author} {\bibfnamefont {R.}~\bibnamefont {Mazur}},
  \bibinfo {author} {\bibfnamefont {M.}~\bibnamefont {Muszy{\'n}ski}}, \bibinfo
  {author} {\bibfnamefont {W.}~\bibnamefont {Piecek}}, \bibinfo {author}
  {\bibfnamefont {P.}~\bibnamefont {Kula}}, \bibinfo {author} {\bibfnamefont
  {T.}~\bibnamefont {Stefaniuk}}, \emph {et~al.},\ }\href@noop {} {\bibfield
  {journal} {\bibinfo  {journal} {Science Advances}\ }\textbf {\bibinfo
  {volume} {8}},\ \bibinfo {pages} {eabq7533} (\bibinfo {year}
  {2022})}\BibitemShut {NoStop}%
\bibitem [{\citenamefont {Li}\ \emph {et~al.}(2022)\citenamefont {Li},
  \citenamefont {Ma}, \citenamefont {Zhai}, \citenamefont {Gao}, \citenamefont
  {Dai}, \citenamefont {Schumacher},\ and\ \citenamefont
  {Gao}}]{li2022manipulating}%
  \BibitemOpen
  \bibfield  {author} {\bibinfo {author} {\bibfnamefont {Y.}~\bibnamefont
  {Li}}, \bibinfo {author} {\bibfnamefont {X.}~\bibnamefont {Ma}}, \bibinfo
  {author} {\bibfnamefont {X.}~\bibnamefont {Zhai}}, \bibinfo {author}
  {\bibfnamefont {M.}~\bibnamefont {Gao}}, \bibinfo {author} {\bibfnamefont
  {H.}~\bibnamefont {Dai}}, \bibinfo {author} {\bibfnamefont {S.}~\bibnamefont
  {Schumacher}},\ and\ \bibinfo {author} {\bibfnamefont {T.}~\bibnamefont
  {Gao}},\ }\href@noop {} {\bibfield  {journal} {\bibinfo  {journal} {Nature
  Communications}\ }\textbf {\bibinfo {volume} {13}},\ \bibinfo {pages} {3785}
  (\bibinfo {year} {2022})}\BibitemShut {NoStop}%
\bibitem [{\citenamefont {Long}\ \emph {et~al.}(2022)\citenamefont {Long},
  \citenamefont {Ma}, \citenamefont {Ren}, \citenamefont {Li}, \citenamefont
  {Liao}, \citenamefont {Schumacher}, \citenamefont {Malpuech}, \citenamefont
  {Solnyshkov},\ and\ \citenamefont {Fu}}]{long2022helical}%
  \BibitemOpen
  \bibfield  {author} {\bibinfo {author} {\bibfnamefont {T.}~\bibnamefont
  {Long}}, \bibinfo {author} {\bibfnamefont {X.}~\bibnamefont {Ma}}, \bibinfo
  {author} {\bibfnamefont {J.}~\bibnamefont {Ren}}, \bibinfo {author}
  {\bibfnamefont {F.}~\bibnamefont {Li}}, \bibinfo {author} {\bibfnamefont
  {Q.}~\bibnamefont {Liao}}, \bibinfo {author} {\bibfnamefont {S.}~\bibnamefont
  {Schumacher}}, \bibinfo {author} {\bibfnamefont {G.}~\bibnamefont
  {Malpuech}}, \bibinfo {author} {\bibfnamefont {D.}~\bibnamefont
  {Solnyshkov}},\ and\ \bibinfo {author} {\bibfnamefont {H.}~\bibnamefont
  {Fu}},\ }\href@noop {} {\bibfield  {journal} {\bibinfo  {journal} {Advanced
  Science}\ }\textbf {\bibinfo {volume} {9}},\ \bibinfo {pages} {2203588}
  (\bibinfo {year} {2022})}\BibitemShut {NoStop}%
\bibitem [{\citenamefont {Liang}\ \emph {et~al.}(2024)\citenamefont {Liang},
  \citenamefont {Wen}, \citenamefont {Jin}, \citenamefont {Rubo}, \citenamefont
  {Liew},\ and\ \citenamefont {Su}}]{liang2024polariton}%
  \BibitemOpen
  \bibfield  {author} {\bibinfo {author} {\bibfnamefont {J.}~\bibnamefont
  {Liang}}, \bibinfo {author} {\bibfnamefont {W.}~\bibnamefont {Wen}}, \bibinfo
  {author} {\bibfnamefont {F.}~\bibnamefont {Jin}}, \bibinfo {author}
  {\bibfnamefont {Y.~G.}\ \bibnamefont {Rubo}}, \bibinfo {author}
  {\bibfnamefont {T.~C.~H.}\ \bibnamefont {Liew}},\ and\ \bibinfo {author}
  {\bibfnamefont {R.}~\bibnamefont {Su}},\ }\href@noop {} {\bibfield  {journal}
  {\bibinfo  {journal} {Nature Photonics}\ }\textbf {\bibinfo {volume} {18}},\
  \bibinfo {pages} {357} (\bibinfo {year} {2024})}\BibitemShut {NoStop}%
\bibitem [{\citenamefont {Muszy{\'n}ski}\ \emph {et~al.}(2024)\citenamefont
  {Muszy{\'n}ski}, \citenamefont {Kokhanchik}, \citenamefont {Urbonas},
  \citenamefont {Kapusci{\'n}ski}, \citenamefont {Oliwa}, \citenamefont
  {Mirek}, \citenamefont {Georgakilas}, \citenamefont {St{\"o}ferle},
  \citenamefont {Mahrt}, \citenamefont {Forster}, \citenamefont {Scherf},
  \citenamefont {Dovzhenko}, \citenamefont {Mazur}, \citenamefont {Morawiak},
  \citenamefont {Piecek}, \citenamefont {Kula}, \citenamefont {Pi\k{e}tka},
  \citenamefont {Solnyshkov}, \citenamefont {Malpuech},\ and\ \citenamefont
  {Szczytko}}]{muszynski2024observation}%
  \BibitemOpen
  \bibfield  {author} {\bibinfo {author} {\bibfnamefont {M.}~\bibnamefont
  {Muszy{\'n}ski}}, \bibinfo {author} {\bibfnamefont {P.}~\bibnamefont
  {Kokhanchik}}, \bibinfo {author} {\bibfnamefont {D.}~\bibnamefont {Urbonas}},
  \bibinfo {author} {\bibfnamefont {P.}~\bibnamefont {Kapusci{\'n}ski}},
  \bibinfo {author} {\bibfnamefont {P.}~\bibnamefont {Oliwa}}, \bibinfo
  {author} {\bibfnamefont {R.}~\bibnamefont {Mirek}}, \bibinfo {author}
  {\bibfnamefont {I.}~\bibnamefont {Georgakilas}}, \bibinfo {author}
  {\bibfnamefont {T.}~\bibnamefont {St{\"o}ferle}}, \bibinfo {author}
  {\bibfnamefont {R.~F.}\ \bibnamefont {Mahrt}}, \bibinfo {author}
  {\bibfnamefont {M.}~\bibnamefont {Forster}}, \bibinfo {author} {\bibfnamefont
  {U.}~\bibnamefont {Scherf}}, \bibinfo {author} {\bibfnamefont
  {D.}~\bibnamefont {Dovzhenko}}, \bibinfo {author} {\bibfnamefont
  {R.}~\bibnamefont {Mazur}}, \bibinfo {author} {\bibfnamefont
  {P.}~\bibnamefont {Morawiak}}, \bibinfo {author} {\bibfnamefont
  {W.}~\bibnamefont {Piecek}}, \bibinfo {author} {\bibfnamefont
  {P.}~\bibnamefont {Kula}}, \bibinfo {author} {\bibfnamefont {B.}~\bibnamefont
  {Pi\k{e}tka}}, \bibinfo {author} {\bibfnamefont {D.}~\bibnamefont
  {Solnyshkov}}, \bibinfo {author} {\bibfnamefont {G.}~\bibnamefont
  {Malpuech}},\ and\ \bibinfo {author} {\bibfnamefont {J.}~\bibnamefont
  {Szczytko}},\ }\href@noop {} {\bibfield  {journal} {\bibinfo  {journal}
  {arXiv preprint arXiv:2407.02406}\ } (\bibinfo {year} {2024})}\BibitemShut
  {NoStop}%
\bibitem [{\citenamefont {Jin}\ \emph {et~al.}(2006)\citenamefont {Jin},
  \citenamefont {Li},\ and\ \citenamefont {Zhang}}]{jin20062}%
  \BibitemOpen
  \bibfield  {author} {\bibinfo {author} {\bibfnamefont {P.-Q.}\ \bibnamefont
  {Jin}}, \bibinfo {author} {\bibfnamefont {Y.-Q.}\ \bibnamefont {Li}},\ and\
  \bibinfo {author} {\bibfnamefont {F.-C.}\ \bibnamefont {Zhang}},\ }\href@noop
  {} {\bibfield  {journal} {\bibinfo  {journal} {Journal of Physics A:
  Mathematical and General}\ }\textbf {\bibinfo {volume} {39}},\ \bibinfo
  {pages} {7115} (\bibinfo {year} {2006})}\BibitemShut {NoStop}%
\bibitem [{\citenamefont {Bernevig}\ \emph
  {et~al.}(2006{\natexlab{a}})\citenamefont {Bernevig}, \citenamefont
  {Orenstein},\ and\ \citenamefont {Zhang}}]{bernevig2006exact}%
  \BibitemOpen
  \bibfield  {author} {\bibinfo {author} {\bibfnamefont {B.~A.}\ \bibnamefont
  {Bernevig}}, \bibinfo {author} {\bibfnamefont {J.}~\bibnamefont
  {Orenstein}},\ and\ \bibinfo {author} {\bibfnamefont {S.-C.}\ \bibnamefont
  {Zhang}},\ }\href@noop {} {\bibfield  {journal} {\bibinfo  {journal}
  {Physical Review Letters}\ }\textbf {\bibinfo {volume} {97}},\ \bibinfo
  {pages} {236601} (\bibinfo {year} {2006}{\natexlab{a}})}\BibitemShut
  {NoStop}%
\bibitem [{\citenamefont {Koralek}\ \emph {et~al.}(2009)\citenamefont
  {Koralek}, \citenamefont {Weber}, \citenamefont {Orenstein}, \citenamefont
  {Bernevig}, \citenamefont {Zhang}, \citenamefont {Mack},\ and\ \citenamefont
  {Awschalom}}]{koralek2009emergence}%
  \BibitemOpen
  \bibfield  {author} {\bibinfo {author} {\bibfnamefont {J.~D.}\ \bibnamefont
  {Koralek}}, \bibinfo {author} {\bibfnamefont {C.~P.}\ \bibnamefont {Weber}},
  \bibinfo {author} {\bibfnamefont {J.}~\bibnamefont {Orenstein}}, \bibinfo
  {author} {\bibfnamefont {B.~A.}\ \bibnamefont {Bernevig}}, \bibinfo {author}
  {\bibfnamefont {S.-C.}\ \bibnamefont {Zhang}}, \bibinfo {author}
  {\bibfnamefont {S.}~\bibnamefont {Mack}},\ and\ \bibinfo {author}
  {\bibfnamefont {D.}~\bibnamefont {Awschalom}},\ }\href@noop {} {\bibfield
  {journal} {\bibinfo  {journal} {Nature}\ }\textbf {\bibinfo {volume} {458}},\
  \bibinfo {pages} {610} (\bibinfo {year} {2009})}\BibitemShut {NoStop}%
\bibitem [{\citenamefont {Tang}\ \emph {et~al.}(2022)\citenamefont {Tang},
  \citenamefont {He}, \citenamefont {Shi}, \citenamefont {Liu}, \citenamefont
  {Chen},\ and\ \citenamefont {Dong}}]{tang2022topological}%
  \BibitemOpen
  \bibfield  {author} {\bibinfo {author} {\bibfnamefont {G.-J.}\ \bibnamefont
  {Tang}}, \bibinfo {author} {\bibfnamefont {X.-T.}\ \bibnamefont {He}},
  \bibinfo {author} {\bibfnamefont {F.-L.}\ \bibnamefont {Shi}}, \bibinfo
  {author} {\bibfnamefont {J.-W.}\ \bibnamefont {Liu}}, \bibinfo {author}
  {\bibfnamefont {X.-D.}\ \bibnamefont {Chen}},\ and\ \bibinfo {author}
  {\bibfnamefont {J.-W.}\ \bibnamefont {Dong}},\ }\href@noop {} {\bibfield
  {journal} {\bibinfo  {journal} {Laser \& Photonics Reviews}\ }\textbf
  {\bibinfo {volume} {16}},\ \bibinfo {pages} {2100300} (\bibinfo {year}
  {2022})}\BibitemShut {NoStop}%
\bibitem [{\citenamefont {Kokhanchik}\ \emph {et~al.}(2022)\citenamefont
  {Kokhanchik}, \citenamefont {Solnyshkov}, \citenamefont {St{\"o}ferle},
  \citenamefont {Pi\k{e}tka}, \citenamefont {Szczytko},\ and\ \citenamefont
  {Malpuech}}]{kokhanchik2022modulated}%
  \BibitemOpen
  \bibfield  {author} {\bibinfo {author} {\bibfnamefont {P.}~\bibnamefont
  {Kokhanchik}}, \bibinfo {author} {\bibfnamefont {D.}~\bibnamefont
  {Solnyshkov}}, \bibinfo {author} {\bibfnamefont {T.}~\bibnamefont
  {St{\"o}ferle}}, \bibinfo {author} {\bibfnamefont {B.}~\bibnamefont
  {Pi\k{e}tka}}, \bibinfo {author} {\bibfnamefont {J.}~\bibnamefont
  {Szczytko}},\ and\ \bibinfo {author} {\bibfnamefont {G.}~\bibnamefont
  {Malpuech}},\ }\href@noop {} {\bibfield  {journal} {\bibinfo  {journal}
  {Physical Review Letters}\ }\textbf {\bibinfo {volume} {129}},\ \bibinfo
  {pages} {246801} (\bibinfo {year} {2022})}\BibitemShut {NoStop}%
\bibitem [{sup()}]{suppl}%
  \BibitemOpen
  \href@noop {} {}\bibinfo {note} {See Supplemental Material.}\BibitemShut
  {Stop}%
\bibitem [{\citenamefont {Bernevig}\ \emph
  {et~al.}(2006{\natexlab{b}})\citenamefont {Bernevig}, \citenamefont
  {Hughes},\ and\ \citenamefont {Zhang}}]{bernevig2006quantum}%
  \BibitemOpen
  \bibfield  {author} {\bibinfo {author} {\bibfnamefont {B.~A.}\ \bibnamefont
  {Bernevig}}, \bibinfo {author} {\bibfnamefont {T.~L.}\ \bibnamefont
  {Hughes}},\ and\ \bibinfo {author} {\bibfnamefont {S.-C.}\ \bibnamefont
  {Zhang}},\ }\href@noop {} {\bibfield  {journal} {\bibinfo  {journal}
  {science}\ }\textbf {\bibinfo {volume} {314}},\ \bibinfo {pages} {1757}
  (\bibinfo {year} {2006}{\natexlab{b}})}\BibitemShut {NoStop}%
\bibitem [{\citenamefont {Michaelis~de Vasconcellos}\ \emph
  {et~al.}(2011)\citenamefont {Michaelis~de Vasconcellos}, \citenamefont
  {Calvar}, \citenamefont {Dousse}, \citenamefont {Suffczy{\'n}ski},
  \citenamefont {Dupuis}, \citenamefont {Lema{\^\i}tre}, \citenamefont
  {Sagnes}, \citenamefont {Bloch}, \citenamefont {Voisin},\ and\ \citenamefont
  {Senellart}}]{michaelis2011spatial}%
  \BibitemOpen
  \bibfield  {author} {\bibinfo {author} {\bibfnamefont {S.}~\bibnamefont
  {Michaelis~de Vasconcellos}}, \bibinfo {author} {\bibfnamefont
  {A.}~\bibnamefont {Calvar}}, \bibinfo {author} {\bibfnamefont
  {A.}~\bibnamefont {Dousse}}, \bibinfo {author} {\bibfnamefont
  {J.}~\bibnamefont {Suffczy{\'n}ski}}, \bibinfo {author} {\bibfnamefont
  {N.}~\bibnamefont {Dupuis}}, \bibinfo {author} {\bibfnamefont
  {A.}~\bibnamefont {Lema{\^\i}tre}}, \bibinfo {author} {\bibfnamefont
  {I.}~\bibnamefont {Sagnes}}, \bibinfo {author} {\bibfnamefont
  {J.}~\bibnamefont {Bloch}}, \bibinfo {author} {\bibfnamefont
  {P.}~\bibnamefont {Voisin}},\ and\ \bibinfo {author} {\bibfnamefont
  {P.}~\bibnamefont {Senellart}},\ }\href@noop {} {\bibfield  {journal}
  {\bibinfo  {journal} {Applied Physics Letters}\ }\textbf {\bibinfo {volume}
  {99}} (\bibinfo {year} {2011})}\BibitemShut {NoStop}%
\bibitem [{\citenamefont {Solnyshkov}\ \emph {et~al.}(2021)\citenamefont
  {Solnyshkov}, \citenamefont {Malpuech}, \citenamefont {St-Jean},
  \citenamefont {Ravets}, \citenamefont {Bloch},\ and\ \citenamefont
  {Amo}}]{solnyshkov2021microcavity}%
  \BibitemOpen
  \bibfield  {author} {\bibinfo {author} {\bibfnamefont {D.~D.}\ \bibnamefont
  {Solnyshkov}}, \bibinfo {author} {\bibfnamefont {G.}~\bibnamefont
  {Malpuech}}, \bibinfo {author} {\bibfnamefont {P.}~\bibnamefont {St-Jean}},
  \bibinfo {author} {\bibfnamefont {S.}~\bibnamefont {Ravets}}, \bibinfo
  {author} {\bibfnamefont {J.}~\bibnamefont {Bloch}},\ and\ \bibinfo {author}
  {\bibfnamefont {A.}~\bibnamefont {Amo}},\ }\href@noop {} {\bibfield
  {journal} {\bibinfo  {journal} {Optical Materials Express}\ }\textbf
  {\bibinfo {volume} {11}},\ \bibinfo {pages} {1119} (\bibinfo {year}
  {2021})}\BibitemShut {NoStop}%
\bibitem [{\citenamefont {Sala}\ \emph {et~al.}(2015)\citenamefont {Sala},
  \citenamefont {Solnyshkov}, \citenamefont {Carusotto}, \citenamefont
  {Jacqmin}, \citenamefont {Lema{\^\i}tre}, \citenamefont {Ter{\c{c}}as},
  \citenamefont {Nalitov}, \citenamefont {Abbarchi}, \citenamefont {Galopin},
  \citenamefont {Sagnes} \emph {et~al.}}]{sala2015spin}%
  \BibitemOpen
  \bibfield  {author} {\bibinfo {author} {\bibfnamefont {V.}~\bibnamefont
  {Sala}}, \bibinfo {author} {\bibfnamefont {D.~D.}\ \bibnamefont
  {Solnyshkov}}, \bibinfo {author} {\bibfnamefont {I.}~\bibnamefont
  {Carusotto}}, \bibinfo {author} {\bibfnamefont {T.}~\bibnamefont {Jacqmin}},
  \bibinfo {author} {\bibfnamefont {A.}~\bibnamefont {Lema{\^\i}tre}}, \bibinfo
  {author} {\bibfnamefont {H.}~\bibnamefont {Ter{\c{c}}as}}, \bibinfo {author}
  {\bibfnamefont {A.}~\bibnamefont {Nalitov}}, \bibinfo {author} {\bibfnamefont
  {M.}~\bibnamefont {Abbarchi}}, \bibinfo {author} {\bibfnamefont
  {E.}~\bibnamefont {Galopin}}, \bibinfo {author} {\bibfnamefont
  {I.}~\bibnamefont {Sagnes}}, \emph {et~al.},\ }\href@noop {} {\bibfield
  {journal} {\bibinfo  {journal} {Physical Review X}\ }\textbf {\bibinfo
  {volume} {5}},\ \bibinfo {pages} {011034} (\bibinfo {year}
  {2015})}\BibitemShut {NoStop}%
\bibitem [{\citenamefont {T{\"o}pfer}\ \emph {et~al.}(2021)\citenamefont
  {T{\"o}pfer}, \citenamefont {Chatzopoulos}, \citenamefont {Sigurdsson},
  \citenamefont {Cookson}, \citenamefont {Rubo},\ and\ \citenamefont
  {Lagoudakis}}]{topfer2021engineering}%
  \BibitemOpen
  \bibfield  {author} {\bibinfo {author} {\bibfnamefont {J.~D.}\ \bibnamefont
  {T{\"o}pfer}}, \bibinfo {author} {\bibfnamefont {I.}~\bibnamefont
  {Chatzopoulos}}, \bibinfo {author} {\bibfnamefont {H.}~\bibnamefont
  {Sigurdsson}}, \bibinfo {author} {\bibfnamefont {T.}~\bibnamefont {Cookson}},
  \bibinfo {author} {\bibfnamefont {Y.~G.}\ \bibnamefont {Rubo}},\ and\
  \bibinfo {author} {\bibfnamefont {P.~G.}\ \bibnamefont {Lagoudakis}},\
  }\href@noop {} {\bibfield  {journal} {\bibinfo  {journal} {Optica}\ }\textbf
  {\bibinfo {volume} {8}},\ \bibinfo {pages} {106} (\bibinfo {year}
  {2021})}\BibitemShut {NoStop}%
\bibitem [{\citenamefont {Proukakis}\ \emph {et~al.}(2017)\citenamefont
  {Proukakis}, \citenamefont {Snoke},\ and\ \citenamefont
  {Littlewood}}]{proukakis2017universal}%
  \BibitemOpen
  \bibfield  {author} {\bibinfo {author} {\bibfnamefont {N.~P.}\ \bibnamefont
  {Proukakis}}, \bibinfo {author} {\bibfnamefont {D.~W.}\ \bibnamefont
  {Snoke}},\ and\ \bibinfo {author} {\bibfnamefont {P.~B.}\ \bibnamefont
  {Littlewood}},\ }\href@noop {} {\emph {\bibinfo {title} {Universal {T}hemes
  of {B}ose-{E}instein {C}ondensation}}}\ (\bibinfo  {publisher} {Cambridge
  {U}niversity {P}ress},\ \bibinfo {year} {2017})\BibitemShut {NoStop}%
\bibitem [{\citenamefont {Lin}\ \emph {et~al.}(2011)\citenamefont {Lin},
  \citenamefont {Jim{\'e}nez-Garc{\'\i}a},\ and\ \citenamefont
  {Spielman}}]{lin2011spin}%
  \BibitemOpen
  \bibfield  {author} {\bibinfo {author} {\bibfnamefont {Y.-J.}\ \bibnamefont
  {Lin}}, \bibinfo {author} {\bibfnamefont {K.}~\bibnamefont
  {Jim{\'e}nez-Garc{\'\i}a}},\ and\ \bibinfo {author} {\bibfnamefont {I.~B.}\
  \bibnamefont {Spielman}},\ }\href@noop {} {\bibfield  {journal} {\bibinfo
  {journal} {Nature}\ }\textbf {\bibinfo {volume} {471}},\ \bibinfo {pages}
  {83} (\bibinfo {year} {2011})}\BibitemShut {NoStop}%
\end{thebibliography}%


\begin{thebibliography}{2}%
\makeatletter
\providecommand \@ifxundefined [1]{%
 \@ifx{#1\undefined}
}%
\providecommand \@ifnum [1]{%
 \ifnum #1\expandafter \@firstoftwo
 \else \expandafter \@secondoftwo
 \fi
}%
\providecommand \@ifx [1]{%
 \ifx #1\expandafter \@firstoftwo
 \else \expandafter \@secondoftwo
 \fi
}%
\providecommand \natexlab [1]{#1}%
\providecommand \enquote  [1]{``#1''}%
\providecommand \bibnamefont  [1]{#1}%
\providecommand \bibfnamefont [1]{#1}%
\providecommand \citenamefont [1]{#1}%
\providecommand \href@noop [0]{\@secondoftwo}%
\providecommand \href [0]{\begingroup \@sanitize@url \@href}%
\providecommand \@href[1]{\@@startlink{#1}\@@href}%
\providecommand \@@href[1]{\endgroup#1\@@endlink}%
\providecommand \@sanitize@url [0]{\catcode `\\12\catcode `\$12\catcode
  `\&12\catcode `\#12\catcode `\^12\catcode `\_12\catcode `\%12\relax}%
\providecommand \@@startlink[1]{}%
\providecommand \@@endlink[0]{}%
\providecommand \url  [0]{\begingroup\@sanitize@url \@url }%
\providecommand \@url [1]{\endgroup\@href {#1}{\urlprefix }}%
\providecommand \urlprefix  [0]{URL }%
\providecommand \Eprint [0]{\href }%
\providecommand \doibase [0]{https://doi.org/}%
\providecommand \selectlanguage [0]{\@gobble}%
\providecommand \bibinfo  [0]{\@secondoftwo}%
\providecommand \bibfield  [0]{\@secondoftwo}%
\providecommand \translation [1]{[#1]}%
\providecommand \BibitemOpen [0]{}%
\providecommand \bibitemStop [0]{}%
\providecommand \bibitemNoStop [0]{.\EOS\space}%
\providecommand \EOS [0]{\spacefactor3000\relax}%
\providecommand \BibitemShut  [1]{\csname bibitem#1\endcsname}%
\let\auto@bib@innerbib\@empty
%</preamble>
\bibitem [{\citenamefont {Mai}\ \emph {et~al.}(2013)\citenamefont {Mai},
  \citenamefont {Ding}, \citenamefont {St\"{o}ferle}, \citenamefont {Knoll},
  \citenamefont {J.~Offrein},\ and\ \citenamefont {Mahrt}}]{Mai2013}%
  \BibitemOpen
  \bibfield  {author} {\bibinfo {author} {\bibfnamefont {L.}~\bibnamefont
  {Mai}}, \bibinfo {author} {\bibfnamefont {F.}~\bibnamefont {Ding}}, \bibinfo
  {author} {\bibfnamefont {T.}~\bibnamefont {St\"{o}ferle}}, \bibinfo {author}
  {\bibfnamefont {A.}~\bibnamefont {Knoll}}, \bibinfo {author} {\bibfnamefont
  {B.}~\bibnamefont {J.~Offrein}},\ and\ \bibinfo {author} {\bibfnamefont
  {R.~F.}\ \bibnamefont {Mahrt}},\ }\href@noop {} {\bibfield  {journal}
  {\bibinfo  {journal} {Applied Physics Letters}\ }\textbf {\bibinfo {volume}
  {103}} (\bibinfo {year} {2013})}\BibitemShut {NoStop}%
\bibitem [{\citenamefont {Rechci{\'n}ska}\ \emph {et~al.}(2019)\citenamefont
  {Rechci{\'n}ska}, \citenamefont {Kr{\'o}l}, \citenamefont {Mazur},
  \citenamefont {Morawiak}, \citenamefont {Mirek}, \citenamefont {{\L}empicka},
  \citenamefont {Bardyszewski}, \citenamefont {Matuszewski}, \citenamefont
  {Kula}, \citenamefont {Piecek} \emph {et~al.}}]{rechcinska2019engineering}%
  \BibitemOpen
  \bibfield  {author} {\bibinfo {author} {\bibfnamefont {K.}~\bibnamefont
  {Rechci{\'n}ska}}, \bibinfo {author} {\bibfnamefont {M.}~\bibnamefont
  {Kr{\'o}l}}, \bibinfo {author} {\bibfnamefont {R.}~\bibnamefont {Mazur}},
  \bibinfo {author} {\bibfnamefont {P.}~\bibnamefont {Morawiak}}, \bibinfo
  {author} {\bibfnamefont {R.}~\bibnamefont {Mirek}}, \bibinfo {author}
  {\bibfnamefont {K.}~\bibnamefont {{\L}empicka}}, \bibinfo {author}
  {\bibfnamefont {W.}~\bibnamefont {Bardyszewski}}, \bibinfo {author}
  {\bibfnamefont {M.}~\bibnamefont {Matuszewski}}, \bibinfo {author}
  {\bibfnamefont {P.}~\bibnamefont {Kula}}, \bibinfo {author} {\bibfnamefont
  {W.}~\bibnamefont {Piecek}}, \emph {et~al.},\ }\href@noop {} {\bibfield
  {journal} {\bibinfo  {journal} {Science}\ }\textbf {\bibinfo {volume}
  {366}},\ \bibinfo {pages} {727} (\bibinfo {year} {2019})}\BibitemShut
  {NoStop}%
\end{thebibliography}%
	
\end{document}

\fancyhf{}
\cfoot{\thepage}
\pagestyle{fancy}  
\pagenumbering{arabic}
\makeatletter
\let\ps@titlepage\ps@plain
\makeatother
    
\title{Supplemental Material\\In-situ tunneling control in photonic potentials by Rashba-Dresselhaus spin-orbit~coupling}

\author{Rafa\l{}\,Mirek}
\altaffiliation[]{These authors contributed equally to this letter.}
\affiliation{IBM Research Europe - Zurich, S{\"a}umerstrasse 4, R{\"u}schlikon, Switzerland}
\email[]{rafal.mirek@ibm.com}

\author{Pavel\,Kokhanchik}
\altaffiliation[]{These authors contributed equally to this letter.}
\affiliation{Universit\'e Clermont Auvergne, Clermont Auvergne INP, CNRS, Institut Pascal, F-63000 Clermont-Ferrand, France}

\author{Darius\,Urbonas}
\affiliation{IBM Research Europe - Zurich, S{\"a}umerstrasse 4, R{\"u}schlikon, Switzerland}

\author{Ioannis\,Georgakilas}
\affiliation{IBM Research Europe - Zurich, S{\"a}umerstrasse 4, R{\"u}schlikon, Switzerland}

\author{Marcin\,Muszy\'nski}
\affiliation{Institute of Experimental Physics, Faculty of Physics, University of Warsaw, Poland}

\author{Piotr\,Kapu\'sci\'nski}
\affiliation{Institute of Experimental Physics, Faculty of Physics, University of Warsaw, Poland}

\author{Przemys\l{}aw\, Oliwa}
\affiliation{Institute of Experimental Physics, Faculty of Physics, University of Warsaw, Poland}

\author{Barbara\,Pi\k{e}tka}
\affiliation{Institute of Experimental Physics, Faculty of Physics, University of Warsaw, Poland}

\author{Jacek\,Szczytko}
\affiliation{Institute of Experimental Physics, Faculty of Physics, University of Warsaw, Poland}

\author{Michael\,Forster}
\affiliation{Macromolecular Chemistry Group and Wuppertal Center for Smart Materials \& Systems (CM@S), Bergische Universit\"{a}t Wuppertal, Gauss Strasse 20, 42119 Wuppertal, Germany}

\author{Ullrich\,Scherf}
\affiliation{Macromolecular Chemistry Group and Wuppertal Center for Smart Materials \& Systems (CM@S), Bergische Universit\"{a}t Wuppertal, Gauss Strasse 20, 42119 Wuppertal, Germany}

\author{Przemys\l{}aw\,Morawiak}
\affiliation{Institute of Applied Physics, Military University of Technology, Warsaw, Poland}

\author{Wiktor\,Piecek}
\affiliation{Institute of Applied Physics, Military University of Technology, Warsaw, Poland}

\author{Przemys\l{}aw\,Kula}
\affiliation{Institute of Chemistry, Military University of Technology, Warsaw, Poland}

\author{Dmitry\,Solnyshkov}
\affiliation{Universit\'e Clermont Auvergne, Clermont Auvergne INP, CNRS, Institut Pascal, F-63000 Clermont-Ferrand, France}
\affiliation{Institut Universitaire de France (IUF), 75231 Paris, France}

\author{Guillaume\,Malpuech}
\affiliation{Universit\'e Clermont Auvergne, Clermont Auvergne INP, CNRS, Institut Pascal, F-63000 Clermont-Ferrand, France}

\author{Rainer\,F.\,Mahrt}
\affiliation{IBM Research Europe - Zurich, S{\"a}umerstrasse 4, R{\"u}schlikon, Switzerland}

\author{Thilo\,St{\"o}ferle}
\affiliation{IBM Research Europe - Zurich, S{\"a}umerstrasse 4, R{\"u}schlikon, Switzerland}

\maketitle
\section{Sample structure}
The sample consists of two dielectric mirrors formed by a distributed Bragg reflector (DBR) deposited on a 3\,mm thick quartz substrate with a 30\,nm thick ITO electrode. Each DBR is made of 7.5~Ta$_2$O$_5$/SiO$_2$ pairs using ion beam deposition. On one of the DBRs, we spin-coated a 40\,nm layer of methyl-substituted ladder-type poly(p-phenylene) (MeLPPP) and covered it with 40\,nm of Al$_2$O$_3$ and 50\,nm of SiO$_2$ spacer. A polyimide layer (DL-3160 by Dalton) was spin-coated on top of the spacer, thermally cured, and then unidirectionally rubbed to serve as an orienting layer for the liquid crystal slab. The second DBR contains artificial Gaussian traps forming ST and DT structures. They were created on the glass substrate with an additional SiO$_2$ layer before the deposition of the mirror using focused ion beam milling~\cite{Mai2013}. The studied ST was 1.53\,$\upmu$m full width at half maximum (FWHM), and the DT potential consisted of two Gaussian traps having FWHM of 1.4\,$\upmu$m and 0.96\,$\upmu$m center-to-center distance. Both structures were etched to a depth of 40\,nm. Finally, both cavity halves were merged and filled with high birefringence ($\Delta$=0.43) liquid crystal 2091 (HBNLC). The spacing and sealing were performed in the same way as in~\cite{rechcinska2019engineering}.

\section{Measurement setup}

\begin{figure*}
    \centering
    \includegraphics[width=.7\linewidth]{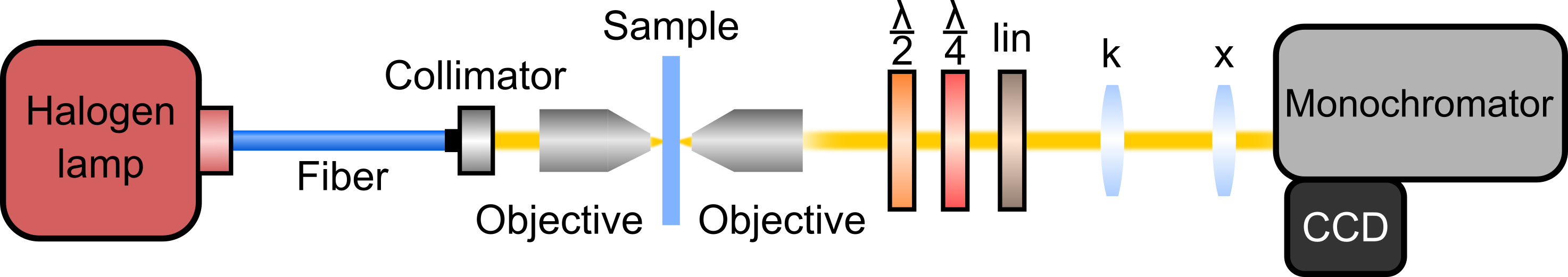}
    \caption{Experimental setup for polarisation-resolved white light transmission experiments.}
    \label{fig_SI_setup}
\end{figure*}

The experimental setup is presented in Fig.~\ref{fig_SI_setup}. White light emitted from the halogen lamp and transmitted through the 10 $\upmu$m diameter fiber (NA = 0.1) is focused on the sample using 50x microscope objective with numerical aperture (NA) = 0.75. The FWHM of the focused beam spot is of around 6 $\upmu$m. The light transmitted through the sample is collected using 10x microscope objective (NA = 0.3) and directed on the imaging spectrograph. We use 0.5 m long monochromator with 300 lines/mm grating. The half-wave plate ($\lambda/2$), quarter-wave plate ($\lambda/4$) and linear polarizer (lin) are used for detecting light of different polarization. The lenses are used for imaging of real space (x) and momentum space (x and k).

\section{Tight-binding model}

The 6x6 tight-binding Hamiltonian describing DT localized states written in the basis $(\ket{H_s^e} \ \ket{H_s^o} \ \ket{V_s^e} \ \ket{V_s^o} \ \ket{H_p^e} \ \ket{V_p^e})^T$ reads as:
\begin{widetext}
    \begin{equation}
        H_{DT}^{(6)} = 
        \begin{pmatrix}
            E_{H_s^e} & 0 & 0 & i \beta_{DT,1} & 0 & 0 \\
            0 & E_{H_s^o} & -i \beta_{DT,1} & 0 & 0& -i \beta_{DT,2} \\
            0 & i \beta_{DT,1} & E_{V_s^e} + \delta & 0 & 0 & 0 \\
            -i \beta_{DT,1} & 0 & 0 & E_{V_s^o} + \delta & -i \beta_{DT,3} & 0 \\
            0 & 0 & 0 & i \beta_{DT,3} & E_{H_p^e} & 0 \\
            0 & i \beta_{DT,2} & 0 & 0 & 0 & E_{V_p^e} + \delta
        \end{pmatrix},
    \label{Ham_DT_6}
    \end{equation}
\end{widetext}
with $E_\varphi$ energy of the localized state $\ket{\varphi}$, $\delta$ linear polarization detuning, $\beta_{DT,1} = |\bra{H_s^e} 2 \alpha k_x \ket{V_s^o}|$, $\beta_{DT,2} = |\bra{H_s^o} 2 \alpha k_x \ket{V_p^e}|$ and $\beta_{DT,3} = |\bra{V_s^o} 2 \alpha k_x \ket{H_p^e}|$ RDSOC matrix elements. The tunneling coefficient between ST localized states $\ket{H_s}$ in the case of zero RDSOC is defined as $j_0=(E_{H_s^o} - E_{H_s^e})/2$, while the degenerate eigenenergies of these states are $E_{H_s}=(E_{H_s^o} + E_{H_s^e})/2$.

After applying 2nd order perturbation theory in the limit $|\delta| \gg \max{|\Delta E_{ij}|}, \beta, |j_0|$, where $\Delta E_{ij}$ is the difference between $i$th and $j$th localized states, we reduce Hamiltonian~\eqref{Ham_DT_6} to an effective 2x2 Hamiltonian including only $(\ket{H_s^l} \ \ket{H_s^r})^T$ basis states (left and right wells in DT):
\begin{equation}
    H_{DT} =
    \begin{pmatrix}
        E_{H_s}'(\delta) & -j(\delta) \\
        -j(\delta) & E_{H_s}'(\delta)
    \end{pmatrix},
\label{Ham_DT_2}
\end{equation}
with $E_{H_s}'(\delta) = E_{H_s} - \beta_{DT,1}^2/(E_{V_s} - E_{H_s} + \delta) - \beta_{DT,2}^2/[2 (E_{V_p^e} - E_{H_s} + \delta)]$ detuning-dependent modified energy of ST state $\ket{H_s}$, and $j(\delta) = j_0 - \beta_{DT,2}^2/[2 (E_{V_p^e} - E_{H_s} + \delta)]$ detuning-dependent effective tunneling. Left-right site basis $(\ket{H_s^l} \ \ket{H_s^r})^T$ is related to the even-odd DT state basis $(\ket{H_s^e} \ \ket{H_s^o})^T$ through the 2x2 Hadamard matrix.

\section{Calculation parameters}
The potentials used for ST and DT simulations are, respectively:
\begin{equation}
    U_{H,V}(x) = -A_{H,V} \exp{\left[-\frac{x^2}{2 \sigma^2} \right]},
\end{equation}
\begin{equation}
   \begin{split}
        & U_{H,V}(x) = -A_{H,V} \times \\ & \left( \exp{\left[-\frac{(x - d/2)^2}{2 \sigma^2} \right]} + \exp{\left[-\frac{(x + d/2)^2}{2 \sigma^2} \right]} \right),
    \end{split} 
\end{equation}
with $A_{H,V}$ potential depth for H,V polarization, $\sigma$ potential width, $d$ intersite distance in DT. The parameters used in the simulation are the following:

Figs.~1-2. Numerical simulation: $\alpha=2$ meV$\cdot \mu$m, $m_H=1.8 \cdot 10^{-5} m_e$, $m_V=2.3 \cdot 10^{-5} m_e$, $\sigma=0.54$ $\mu$m, $A_H=36.64$ meV, $A_V=9.41$ meV. Tight-binding: $E_{H_s} = -25.8$ meV, $E_{H_p} = -9.5$ meV, $E_{V_s} = -5.2$ meV, $E_{V_p} = 1.5$ meV, $\beta_{ST,1} = 7$ meV, $\beta_{ST,2} = 5$ meV.

Fig.~3. Numerical simulation: $\alpha=2.4$ meV$\cdot \mu$m, $m_H=1.8 \cdot 10^{-5} m_e$, $m_V=2.3 \cdot 10^{-5} m_e$, $\sigma=0.45$ $\mu$m, $A_H=32.50$ meV, $A_V=17.73$ meV, $d=1.17$ $\mu$m. Tight-binding: $j_0 = 4.28$ meV, $\beta_{DT,2} = 6.86$ meV, $E_{V_p^e} - E_{H_s} = 7.54$ meV.

Fig.~4. $\delta=20$ meV, $m_H=1.8 \cdot 10^{-5} m_e$, $m_V=2.3 \cdot 10^{-5} m_e$, $\sigma=0.38$ $\mu$m, $A_H=32.50$ meV, $A_V=17.73$ meV, $d=1.30$ $\mu$m.

\begin{figure}
    \centering
    \includegraphics[width=\linewidth]{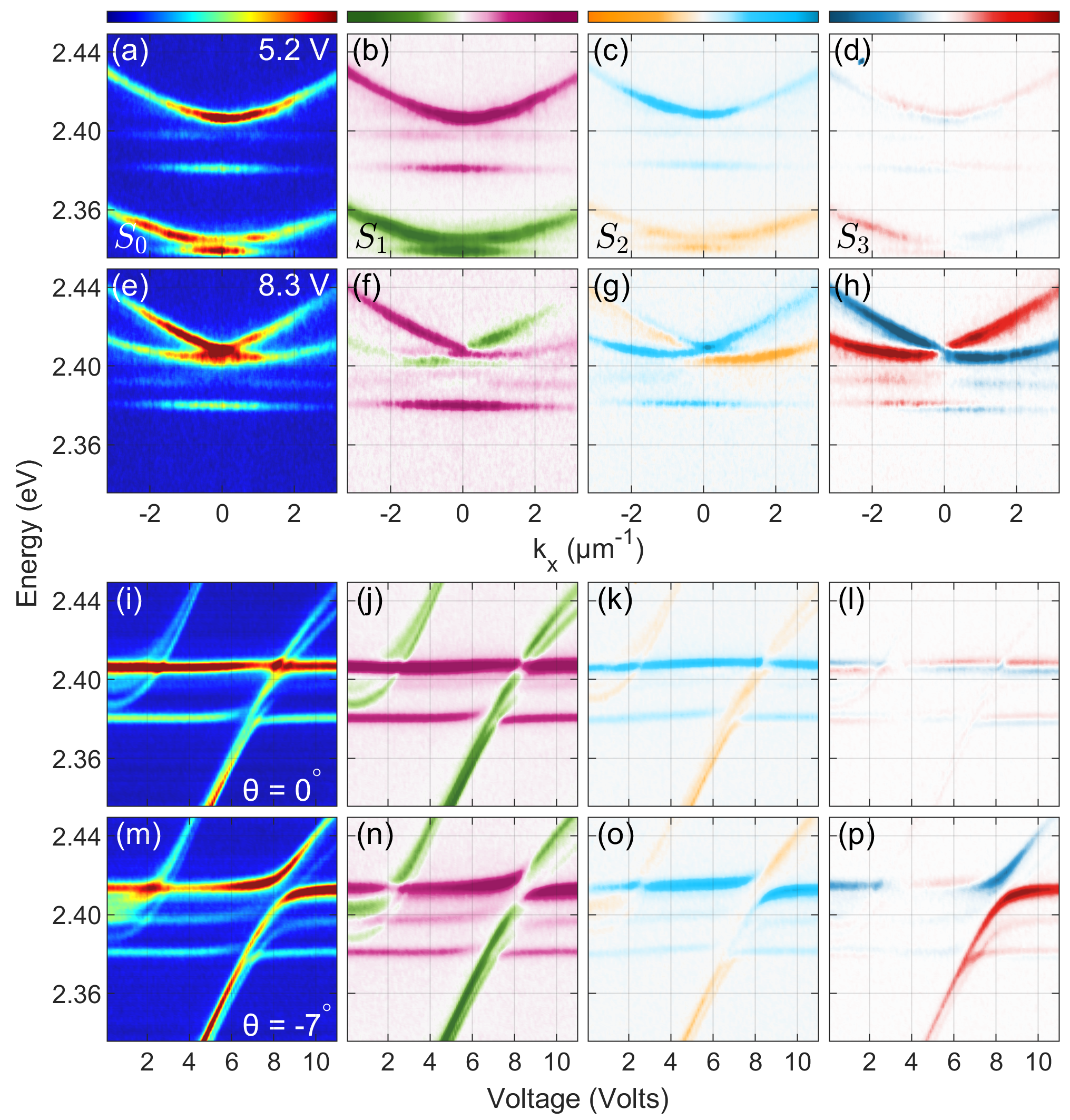}
    \caption{\textbf{Stokes parameters of ST.} Columns (from left to right): Intensity $S_0$ and Stokes parameters $S_{1-3}$; Rows (from top to bottom): (a-d) dispersion for 5.2~V (big linear detuning $\delta$), (e-h) dispersion for 8.3~V (small linear detuning $\delta$), (i-l) energy-voltage diagram for detection angle $\theta = 0^\circ$, (m-p) energy-voltage diagram for detection angle $\theta = -7^\circ$.}
    \label{fig_sm_2}
\end{figure}

\begin{figure}
    \centering
    \includegraphics[width=\linewidth]{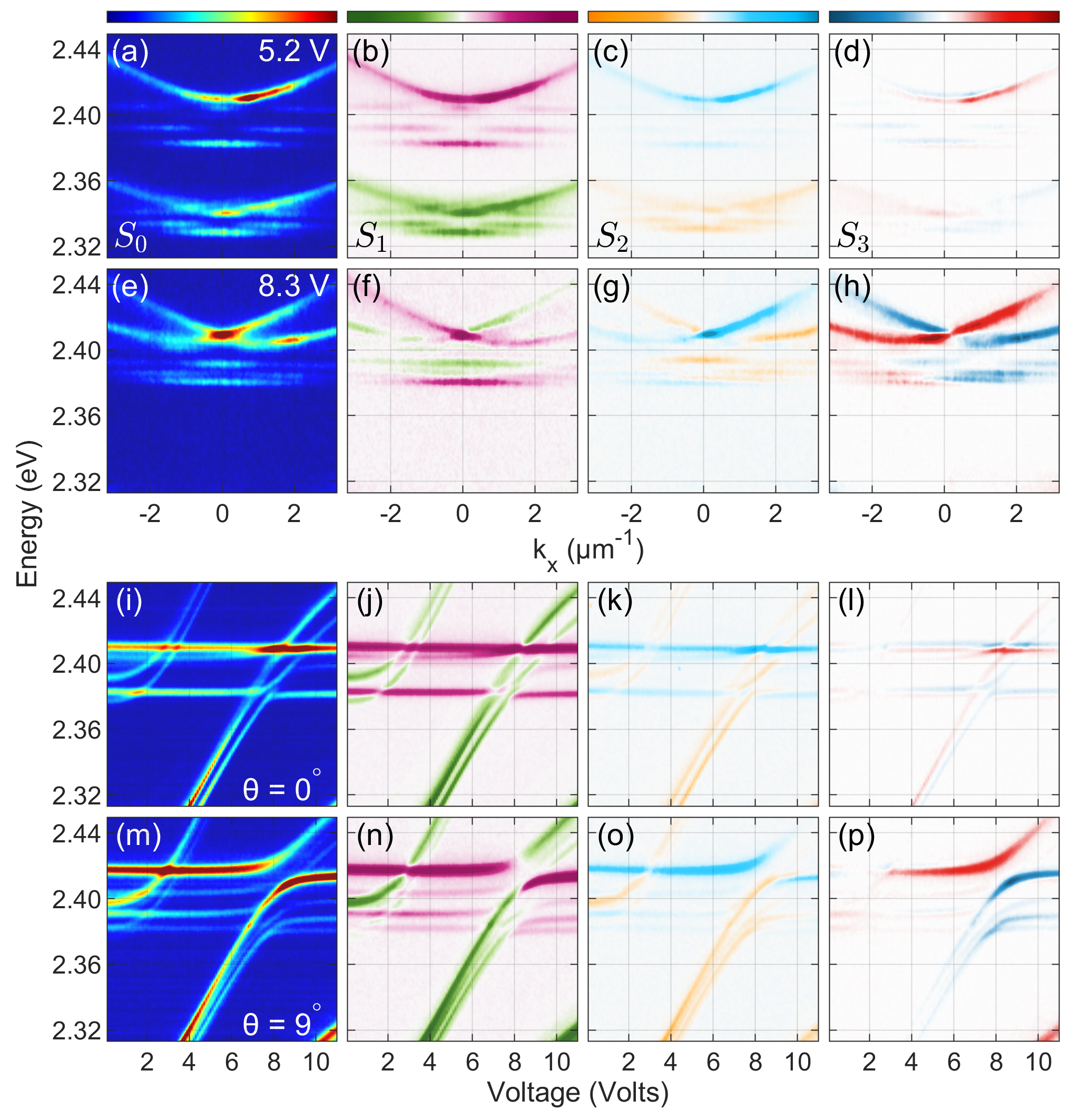}
    \caption{\textbf{Stokes parameters of DT.} Columns (from left to right): Intensity $S_0$ and Stokes parameters $S_{1-3}$; Rows (from top to bottom): (a-d) dispersion for 5.2~V (big linear detuning $\delta$), (e-h) dispersion for 8.3~V (small linear detuning $\delta$), (i-l) energy-voltage diagram for detection angle $\theta = 0^\circ$, (m-p) energy-voltage diagram for detection angle $\theta = 9^\circ$.}
    \label{fig_sm_3}
\end{figure}

\bibliography{si_arxiv}